\documentclass[fleqn,usenatbib]{mnras}
\usepackage{mathptmx}
\usepackage[varvw]{newtxmath}  
\usepackage[T1]{fontenc}
\usepackage{ae,aecompl}
\usepackage{times}
\usepackage{amsmath}
\usepackage{upgreek}
\usepackage{xcolor}
\usepackage{graphicx}
\usepackage{pdflscape}
\usepackage{multicol} 
\usepackage{amsmath,bm}
\usepackage{caption}

\newcommand{\dif}{\mathrm{d}}

\title[From inflow to infall in hub-filament systems]{ATOMS: ALMA Three-millimeter Observations of Massive Star-forming regions -- XI. From inflow to infall in hub-filament systems}
\author[J. W. Zhou, T. Liu et al.]{
Jian-Wen Zhou,\thanks{E-mail: zjw@nao.cas.cn }$^{1,2}$
Tie Liu,\thanks{E-mail: liutie@shao.ac.cn }$^{3}$
Neal J. Evans II,$^{4}$
Guido Garay,$^{5}$
Paul F. Goldsmith,$^{6}$
Gilberto C. Gómez,$^{9}$
\newauthor
Enrique Vázquez-Semadeni,$^{9}$ 
Hong-Li Liu,$^{8}$
Amelia M.\ Stutz,$^{13}$
Ke Wang,$^{16}$
Mika Juvela,$^{7}$
Jinhua He,$^{10,11,12}$
\newauthor
Di Li,$^{1,2,15}$
Leonardo Bronfman,$^{12}$
Xunchuan Liu,$^{3}$
Feng-Wei Xu,$^{16,17}$
Anandmayee Tej,$^{28}$
L. K. Dewangan,$^{27}$
\newauthor
Shanghuo Li,$^{29}$
Siju Zhang,$^{16}$
Chao Zhang,$^{19}$
Zhiyuan Ren,$^{1}$
Ken'ichi Tatematsu,$^{20,21}$
Pak Shing Li,$^{26}$
\newauthor
Chang Won Lee,$^{29,33}$
Tapas Baug,$^{22}$
Sheng-Li Qin,$^{8}$
Yuefang Wu,$^{17}$
Yaping Peng,$^{18}$
Yong Zhang,$^{23,24,25}$
\newauthor
Rong Liu,$^{1,2}$
Qiu-Yi Luo,$^{3,2}$
Jixing Ge,$^{11}$
Anindya Saha,$^{28}$
Eswaraiah Chakali,$^{31}$
Qizhou zhang$^{34}$,
\newauthor 
Kee-Tae Kim$^{29}$, 
Isabelle Ristorcelli$^{35}$, 
Zhi-Qiang Shen$^{3}$,
Jin-Zeng Li,\thanks{E-mail: ljz@nao.cas.cn}$^{1}$
\\
Affiliations are listed at the end of the paper}

\date{Accepted XXX. Received YYY; in original form ZZZ}
\pubyear{2021}
\begin{document}
\label{firstpage}
\maketitle

\begin{abstract}
We investigate the presence of hub-filament systems in a large sample of 146 active proto-clusters, using H$^{13}$CO$^{+}$ J=1-0 molecular line data obtained from the ATOMS survey. We find that filaments are  ubiquitous in proto-clusters, and hub-filament systems are very common from dense core scales ($\sim$0.1 pc) to clump/cloud scales ($\sim$1-10 pc). The proportion of proto-clusters containing hub-filament systems decreases with increasing dust temperature ($T_d$) and luminosity-to-mass ratios ($L/M$) of clumps, indicating that stellar feedback from H{\sc ii} regions gradually destroys the hub-filament systems as proto-clusters evolve. Clear velocity gradients are seen along the longest filaments with a mean velocity gradient of 8.71 km s$^{-1}$pc$^{-1}$ and a median velocity gradient of 5.54 km s$^{-1}$pc$^{-1}$. We find that velocity gradients are small for filament lengths larger than $\sim$1~pc, probably hinting at the existence of inertial inflows, although we cannot determine whether the latter are driven by large-scale turbulence or large-scale gravitational contraction. In contrast, velocity gradients below $\sim$1~pc dramatically increase as filament lengths decrease, indicating that the gravity of the hubs or cores starts to dominate gas infall at small scales. We suggest that self-similar hub-filament systems and filamentary accretion at all scales may play a key role in high-mass star formation.
\end{abstract}


\section{Introduction} \label{introduction}

Filamentary structures are ubiquitous in high-mass star-forming molecular clouds. Their relation with massive star formation is not yet understood. According to previous studies, a collapsing cloud with converging filaments forms a hub-filament system. Hub-filament systems are known as a junction of three or more filaments. Filaments have lower column densities compared to the hubs, but filaments show much higher aspect ratio than the hubs \citep{Myers2009,Schneider2012}. In such systems, converging flows are funneling matter into the hub through the filaments. Many case studies have suggested that hub-filament systems are birth cradles of 
high-mass stars and clusters \citep{Peretto2013,Henshaw2014,Zhang2015,Liu2016,Yuan2018,Lu2018,Liuh2019,Issac2019,Dewangan2020}. 
In hub-filament systems, cores embedded in denser clumps can prolong the accretion time for growing massive stars due to the sustained supply of matter from the filamentary environment \citep{Myers2009}. Numerical simulations of colliding flows and collapsing turbulent clumps show that massive protostars grow from low-mass stellar seeds by feeding gas along the dense filamentary streams converging toward the $\sim$0.1 pc size hubs with detectable velocity gradients along the filaments \citep{Wang2010,Gomez2014,Smith2016,Padoan2020}. To date, however, only a few spectral line observations have been conducted to investigate accretion flows along filaments \citep{Liu2012,Kirk2013,Peretto2013,Lu2018,Liuh2019,Chen2019,Chung2019-877,Chung2021-919}. 

To deepen our understanding of high-mass star formation in hub-filament systems, it is necessary to obtain kinematic information on scales down to $\sim$0.1 pc, similar to studies in nearby clouds \citep{Andre2014,Andre2016}. Such high spatial resolution observations toward massive filaments, however, are still rare, and most previous studies are case studies toward infrared dark clouds (IRDCs) \citep{Wang2011,Peretto2013,Henshaw2014,Zhang2015,Beuther2015,Busquet2016,Ohashi2016,Lu2018,Xie2021,Liu2021arXiv211102231L,Liu2022b,2021arXiv211112593L}. Therefore, it is crucial to study the properties of hub-filament systems and investigate how these systems evolve from a statistical view with a large sample.

To this end, here we conduct a statistical study of a large hub-filament sample across various evolutionary phases for investigating the relation between hub-filaments and high-mass star formation using the H$^{13}$CO$+$ J=1-0 molecular line data from ALMA Three-millimeter Observations of Massive Star-forming regions (ATOMS) survey \citep{Liu2020}. We describe the
detailed observations and data used in this paper in section \ref{sec:obs}. The identification and properties of hub-filaments are presented in
Section \ref{sec:results}, and the role of hub-filament systems in high-mass star formation is
discussed in Section \ref{sec:discussion}. We summarize our results in Section \ref{sec:summary}.

\section{Observations} 
\label{sec:obs}

\subsection{ALMA observations}
\label{alma}

We use ALMA data from the ATOMS survey (Project ID: 2019.1.00685.S; PI: Tie Liu). The details of the 12m array and 7m array ALMA observations were summarised in \cite{Liu2020,Liuh2021}. Calibration and imaging were carried out using the CASA software package version 5.6 \citep{McMullin2007}. The 7m data and 12m array data were calibrated separately. Then the visibility data from the 7m and 12m array configurations were combined and later imaged in CASA. For each sourc and each spectral window (spw), a line-free frequency range is automatically determined using the ALMA pipeline (see ALMA technical handbook). This frequency range is used to (a) subtract continuum from line emission in the visibility domain, and (b) make continuum images. Continuum images are made from multi-frequency synthesis of data in this line-free frequency ranges in the two 1.875 GHz wide spectral windows, spw 7 and 8, centered on $\sim99.4$ GHz (or 3 mm). Visibility data from the 12m and 7m array are jointly cleaned using task tclean in CASA 5.6. We used natural weighting and a multiscale deconvolver, for an optimized sensitivity and image quality. All images are primary-beam corrected. The continuum image reaches a typical 1 $\sigma$ rms noise of $\sim$0.2 mJy in a synthesized beam FWHM size of $\sim2.2\arcsec$. 
In this work, we also use H$^{13}$CO$^+$ J=1-0 (86.754288 GHz) line data with a spectral resolution of 0.211 km~s$^{-1}$. The typical beam FWHM size and channel rms noise level for H$^{13}$CO$^+$ J=1-0 line emission are $\sim2.5\arcsec$ and 8 mJy~beam$^{-1}$, respectively. The typical maximum recovered angular scale in this survey is about 1 arcmin, which is comparable to the size of field of view (FOV) in 12-m array observations \citep{Liu2020}. Therefore, missing flux should not be a big issue for gas kinematics studies in this work, even though total-power observations were not included.

\subsection{Spitzer infrared data }
We also use images at 3.6, 4.5, 5.8, and 8.0 $\mu$m, obtained by the \emph{Spitzer} Infrared Array Camera (IRAC), as part of the GLIMPSE project \citep{Benjamin2003}. The images of IRAC were retrieved from the \emph{Spitzer} Archive and the angular resolutions of images are better than $2\arcsec$.

\subsection{Numerical simulation data}

We compared the observational data to one of the filaments described in \citet{Gomez2014}.
Those authors studied the structure and dynamics of filaments formed in a molecular cloud modeled using the {\sc gadget-2} SPH code modified to include the cooling function proposed by \citet{KoyamaInutsuka02} (as corrected for typographical errors by \citealt{Vazquez2007}).
The simulation setup is a high resolution version of the one presented in \citet{Vazquez2007}, which consisted of two transonic streams moving in opposite directions in a medium of constant density (1 cm$^{-3}$) corresponding to the warm neutral medium, in a numerical box of 256 pc per side. The high-resolution simulation of \citet{Gomez2014} had $296^3$ SPH particles, each with a mass of $0.02\, \text{M}_\odot$, and a total mass of $5.18 \times 10^5\, \text{M}_\odot$.
The compression generated by these converging flows induces a phase transition in the gas to the cold neutral medium, so that the Jeans mass of the dense layer decreases by a factor of $10^4$ and its subsequent gravitational collapse is nearly pressureless.
The layer experiences hydrodynamical instabilities 
\citep{Vishniac94,WalderFollini00,Heitsch+05,Heitsch+06,Vazquez+06}
due to the ram-pressure confinement exerted by the flow, causing moderately supersonic turbulent motions within the layer.
Additionally, this flow increases the layer's mass as it collapses, eventually reaching $> 5 \times 10^4 \,\text{M}_\odot$.

Pressureless gravitational collapse of spheroids proceeds along their shortest dimension first \citep{Lin+65}.
Thus, three-dimensional structures collapse into sheets and these in turn collapse into filaments. Therefore, the filaments themselves in this simulation are produced by gravity-driven motions.
\citet{Gomez2014} report filaments of $\sim 15 \,\text{pc}$ and $\sim 600 \,\text{M}_\odot$ when a density threshold of $10^3 \,\text{cm}^{-3}$ is used to define them.
These filaments are not material, but \emph{flow structures} able to exist for times longer than their flow-crossing time, since they are continuously replenished by their surrounding, lower-density cloud gas.
They are able to reach a quasi-steady state because the accreted material is evacuated along their longitudinal direction, onto the clumps that are formed within the filaments or at the positions where two or more filaments meet.
These authors interpret this longitudinal flow ($\sim 0.5 \,\text{km}\,\text{s}^{-1}$ in their simulations) as a signature of the global hierarchical  collapse model for molecular clouds \citep{Vazquez2019}.
The simulation setup (compression-induced phase transition in the neutral medium) ensures that the turbulent flow generated by the instabilities and gravitational collapse, together with the resulting nonlinear density fluctuations, are fully self-consistent, thus avoiding the possibility of artificially supporting the cloud against collapse due to over-driven turbulence injection or spatially mismatched velocity and density structures.
This self-consistency is fundamental for the study of the flow into and along dense filaments.

\section{Results}
\label{sec:results}

\subsection{Filaments identified by \texorpdfstring{$H^{13}CO^{+}$}~J=1-0}\label{identify}
\subsubsection{H$^{13}$CO$^{+}$~J=1-0 as a probe for dense gas}\label{tracer}

\begin{figure*}
  \centering
  \includegraphics[width=1\textwidth]{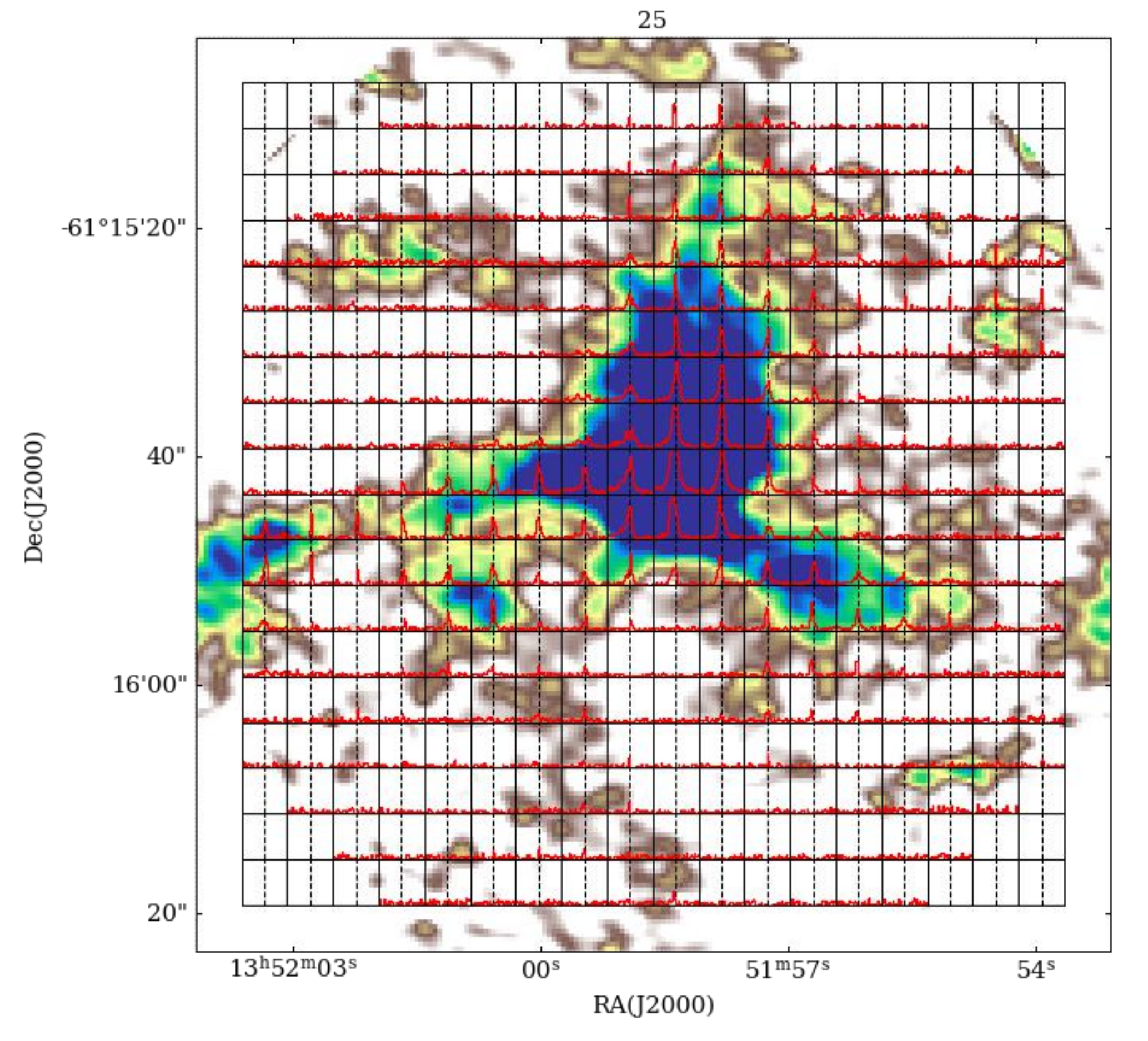}
\caption{The H$^{13}$CO$^+$ spectral grid map for an exemplar source. The background image shows the integrated intensity map. The grid maps of H$^{13}$CO$^+$ for all of ATOMS sources are available as supplementary material.}
\label{grid}
\end{figure*}

\begin{figure*}
  \centering
  \includegraphics[width=1\textwidth]{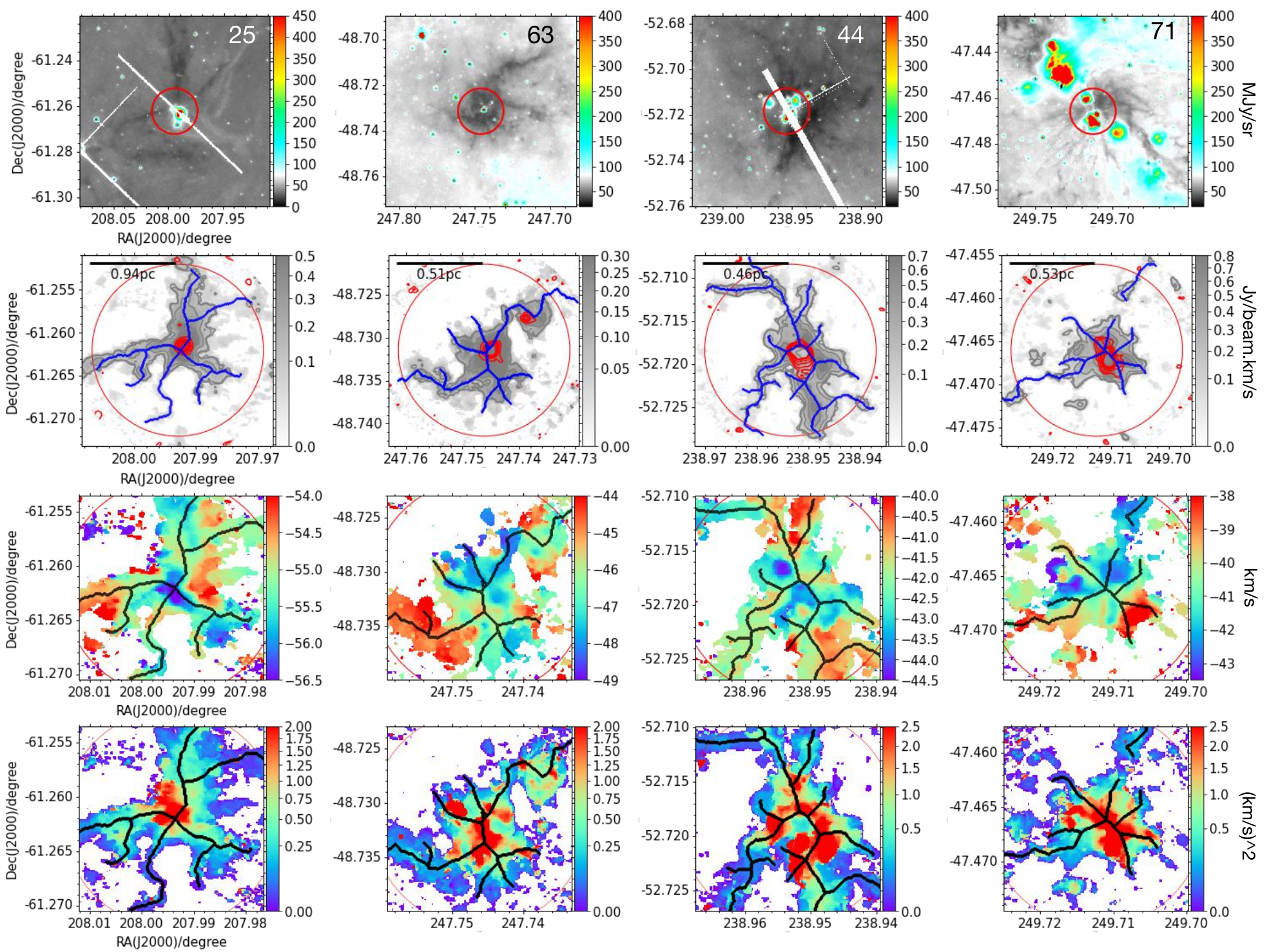}
\caption{Hub-filament systems in four exemplar sources. The first row shows the colour images of \emph{Spitzer} 8 $\mu$m emission.  The source numbers in the upper-right corners are from the Table A1 of \citet{Liu2020}. The second row displays the filaments (blue lines) identified in the moment 0 maps (background) of H$^{13}$CO$^{+}$ J=1-0. The red contours represent 3 mm continuum emission, with lowest contour 5$\sigma$. The third and fourth rows show the moment 1 and 2 maps of H$^{13}$CO$^{+}$ J=1-0 with filament skeletons shown in black lines. The radius of red circles in the maps is 36$\arcsec$, which corresponds to the FOV of the ALMA observations. The 
moment 0 maps of H$^{13}$CO$^{+}$ J=1-0 overlaid with filament skeletons for the other sources are available as supplementary material.}
\label{hf25}
\end{figure*}

We use H$^{13}$CO$^{+}$ J=1-0 line data to identify filaments in the ATOMS sources. H$^{13}$CO$^{+}$ J=1-0 has a rather high critical density of $6.2\times10^{4}$cm$^{-3}$ at 10 K. 
Because its optical depth is much lower than its main line counterpart HCO$^{+}$ J=1-0, its
effective excitation density
is not much less ($n_{\rm eff} = 3.9\times10^{4}$cm$^{-3}$)
\citep{Bergin2007,Shirley2015}. 
These features make it a good tracer of dense gas. Moreover,
\citet{Shimajiri2017} found that the spatial distribution of the H$^{13}$CO$^{+}$ (1-0) emission is tightly correlated with the
column density of the dense gas revealed by \emph{Herschel} data. In particular, the H$^{13}$CO$^{+}$ J=1-0 emission traces the
dense “supercritical” filaments detected by \emph{Herschel} very well \citep{Shimajiri2017}. The virial mass estimates derived from the velocity dispersion of H$^{13}$CO$^{+}$ J=1-0 also agree well with the dense gas mass estimates derived from \emph{Herschel} data for the same sub-regions \citep{Shimajiri2017}. In particular, the H$^{13}$CO$^{+}$ J=1-0 spectra over the mapping areas (except the very dense regions) of the majority ($\sim$75\%) of sources in the ATOMS sample are singly peaked (see Fig.~\ref{grid}). As shown in Fig.~\ref{hf25}, we find that H$^{13}$CO$^{+}$ J=1-0 ATOMS data  trace the whole morphology of hub-filaments well.

\subsubsection{Identification of filaments}\label{filaments}

We use the Moment 0 maps (integrated intensity maps) rather than channel maps of H$^{13}$CO$^{+}$ J=1-0 to identify filaments in ATOMS sources because: (1) ATOMS sources do not show multiple velocity components with velocity differences larger than 10 km~s$^{-1}$ for dense gas tracers (see Fig.~\ref{grid}), which would be attributed to foreground or background cloud emission \citep{Liu2016-829}; (2) Integrated intensity maps have much better sensitivity for identifying fainter gas structures than channel maps; (3) We are interested in the overall hub-filament structures rather than their internal fine structures, 
which may not always be real density enhancements in space due to the complex velocity field \citep{Zamora2017} and are also not well resolved in our observations.

The velocity intervals for making Moment 0 maps are determined from the averaged spectra (radius $\sim25\arcsec$) of H$^{13}$CO$^{+}$ J=1-0, where the intensity of averaged spectra decreases to zero. This limits the typical velocity range within $\sim\pm$5 km~s$^{-1}$ around the systemic velocity for most sources. Moreover, a threshold of 5$\sigma$ is applied to make the moment maps, which can reduce the noise contamination effectively. 

We used the FILFINDER algorithm \citep{Koch2015} to identify filaments from the moment 0 maps of H$^{13}$CO$^{+}$ (1-0). We set the same parameters of FILFINDER for all sources in the first run. However, since the signal-to-noise levels of H$^{13}$CO$^{+}$ J=1-0 emission vary within different sources, we carefully adjust the parameters for individual sources in further identification. We note that the structures identified by FILFINDER change only slightly when we adjust the parameters. The skeletons of identified filaments overlaid on the moment-0 maps of H$^{13}$CO$^{+}$ J=1-0 line emission are shown in Fig.~\ref{hf25} and in supplementary material. The filament skeletons identified by FILFINDER are highly consistent with the gas structures traced by H$^{13}$CO$^{+}$ J=1-0 as seen by eye, indicating that the structures identified in FILFINDER are very reliable. At a first glance of these maps, filaments are nearly ubiquitous in 139 ATOMS proto-clusters. Filaments are not seen in only six ATOMS sources; in all six, the  H$^{13}$CO$^{+}$ J=1-0 line emission was weak or undetected. These six sources may not contain dense gas with densities high enough to excite H$^{13}$CO$^{+}$ J=1-0 line emission.

\subsection{Hub-filament systems in massive proto-clusters}\label{hubfilaments}

\begin{figure*}
  \centering
  \includegraphics[width=1\textwidth]{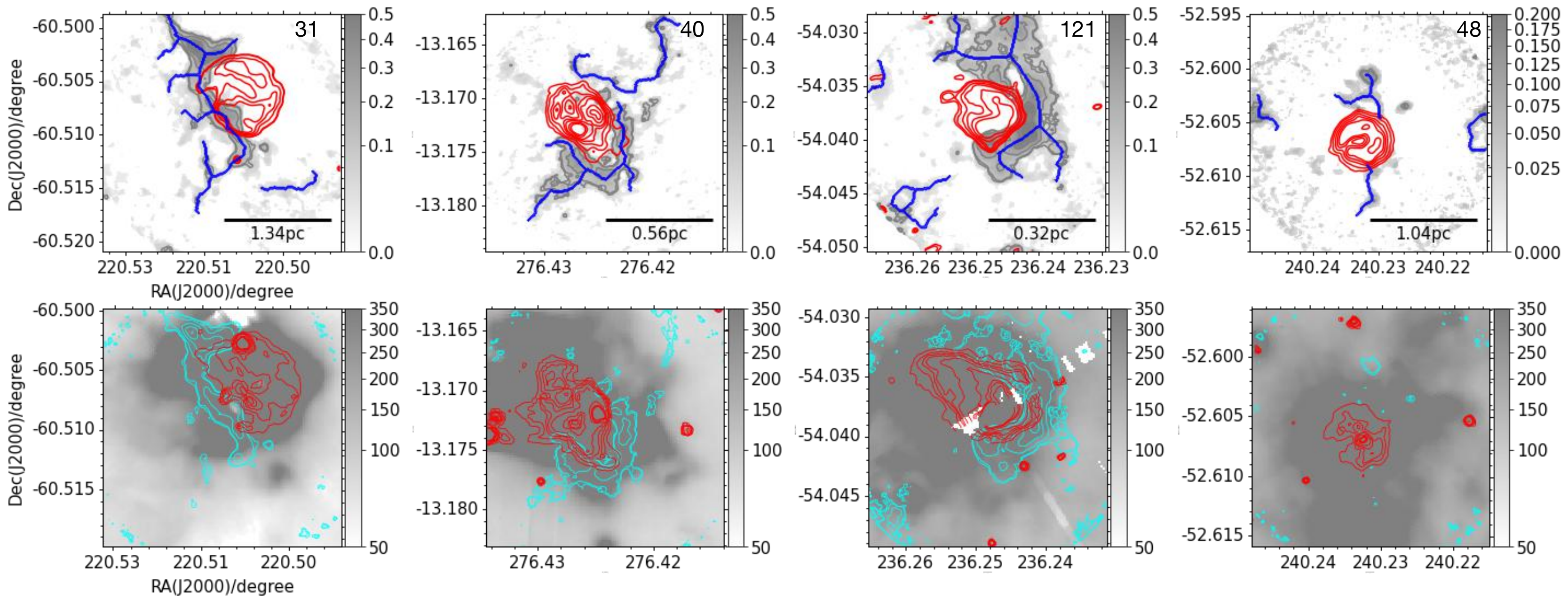}
\caption{Four ATOMS sources containing expanding H{\sc ii} regions. The first row displays the filaments (blue lines) identified in the moment 0 maps (background) of H$^{13}$CO$^{+}$ (1-0). Red contours represent 3mm continuum emission with its minimum contour level of 5$\sigma$. Background images in the second row show the \emph{Spitzer} 8 $\mu$m emission, which traces the PAH emission in PDRs of H{\sc ii} regions. The cyan and red contours show the moment 0 maps of H$^{13}$CO$^{+}$ (1-0) and the \emph{Spitzer} 4.5 $\mu$m emission, respectively. The color bars on the right indicate the flux scale in Jy beam$^{-1}$ km s$^{-1}$ for moment 0 maps and MJy sr$^{-1}$ for \emph{Spitzer} 8 $\mu$m emission. The
moment 0 maps of H$^{13}$CO$^{+}$ J=1-0 overlaid with filament skeletons for the other sources containing H{\sc ii} regions are available as supplementary material.}
\label{hi31}
\end{figure*}

In this work, a strictly defined hub-filament system has following properties: (1) At least three filaments are intertwined; 
(2) The brightest 3 mm continuum core is penetrated by the longest filament, and is close to (less than 1 beam size) the junction region, called the hub, where is the center of the gravity potential well and potential site for high-mass star formation. 

The brightest 3 mm continuum cores in hub regions could be associated with compact H{\sc ii} regions or even younger (Ultra-compact and Hyper-compact) H{\sc ii} regions or high-mass protostars, which are still deeply embedded in molecular gas (see Fig.~\ref{hf25}). Clumps associated with more evolved and extended H{\sc ii} regions that look like infrared bubbles in Spitzer 8 $\mu m$ emission maps (see Fig.~\ref{hi31}) are excluded in the classification of hub-filament systems because these regions are more likely in expansion and are not gravitationally bound. In addition, their 3 mm continuum emission usually significantly deviates from the H$^{13}$CO$^+$ emission. 

Fig.~\ref{hf25} presents the maps of four well-defined hub-filament systems. These hub-filament systems are connected to larger filamentary structures 
revealed by extinction in \emph{Spitzer} 8 $\mu$m images (see top row of panels in Fig.~\ref{hf25}). Although their velocity fields are complicated, most hub-filament systems show clear velocity gradients as shown in the Moment 1 maps of H$^{13}$CO$^{+}$ J=1-0 (third row of Fig.~\ref{hf25}). In addition, the velocity dispersion of H$^{13}$CO$^{+}$ J=1-0 increases toward the hub regions (see bottom panels of Fig.~\ref{hf25}).  

In total, 49$\%$ of the sources are classified as hub-filament systems under our strict definition. This proportion is quite high, considering that the hub-filament morphology of many sources may have been destroyed by stellar feedback. As discussed in \citet{Zhang2021}, nearly 51$\%$ of ATOMS sources show H$_{40\alpha}$ emission, a tracer for H{\sc ii} regions. The hub-filament systems are likely destroyed quickly as H{\sc ii} regions expand (see sources in Fig.~\ref{hi31} for example). The evolution of hub-filament systems under stellar feedback will be discussed in detail below.  

We note that our definition of hub-filament systems is somehow very strict. If we only consider the first condition in the definition and ignore the second one, the fraction of hub-filament systems in ATOMS sources can reach as high as 80\%. To conclude, we find that hub-filament systems are very common within massive proto-clusters.

\subsubsection{Hub-filament systems at various spatial scales}\label{Ubiquitous}

\begin{figure}
\centering
\includegraphics[scale=0.9]{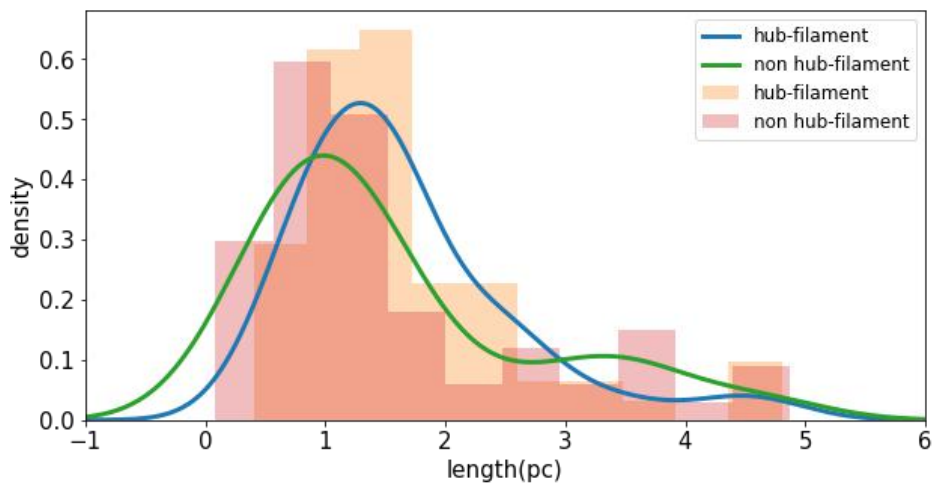}
\caption{The length distribution of the longest filaments in ATOMS sources.}
\label{diameter}
\end{figure}

ATOMS sources are at various distances from 0.4 to 13.0 kpc (see Fig.~\ref{distance}a), enabling us to investigate filaments from core scale ($\sim$0.1 parsec) to clump/cloud scale (several parsec) under a uniform angular resolution ($\sim2\arcsec$) with a large sample \citep{Liu2020}. Fig.~\ref{diameter} shows the length distribution of the longest filaments in these sources. The length ranges from 0.086 pc to 4.87 pc with a mean value of 1.61 pc and a median value of 1.35 pc, indicating that filaments as well as hub-filament systems can exist not only in small-scale ($\sim$0.1 pc) dense cores but also in large-scale clumps/clouds ($\sim$1-5 pc),
and probably up to 10 pc, considering projection effects and the limited FOV of ALMA.

We have also noticed that there is no significant difference in the distribution of filament lengths between our strictly defined hub-filament systems (HFS) sample and the rest non-HFS sample. The possible reasons for this may be: (1) Those non-HFS sources may also contain some "hub"-like structures but cannot match our strict definition of HFS. (2) Those non-HFS sources may have had hubs in the past but their hubs have been destroyed by formed H{\sc ii} regions (cf. Sec.~\ref{evolution}). However, filaments themselves are still persistent. (3) We cannot rule out the possibility that those HFS sources are superpositions of filaments due to projection effect. 

The ATOMS sources were initially selected from a complete and homogeneous CS J=2-1 molecular line survey toward IRAS sources with far-infrared colors characteristics of UC H{\sc ii} regions \citep{Bronfman1996}. As discussed in \citet{Liu2020}, the ATOMS sources are distributed in very different environments of the Milky Way, and are an unbiased sample of the  proto-clusters with the strongest CS J=2-1 line emission (T$_A>2$ K) located in the inner Galatic plane of $-80$\degr$<l<40$\degr, $|b|<2$\degr \citep{Faundez2004-426}. As discussed below, we think that most massive clumps should have had hub-filament systems but their hubs can be gradually destroyed as H{\sc ii} regions form and evolve. Therefore, we suggest that self-similar hub-filament systems at scales from $\sim$0.1 pc to $\sim$10 pc play a crucial role in massive cluster formation in various environments.

\subsubsection{Evolution of Hub-filament systems}\label{evolution}

\begin{figure}
  \centering
  \includegraphics[width=0.45\textwidth]{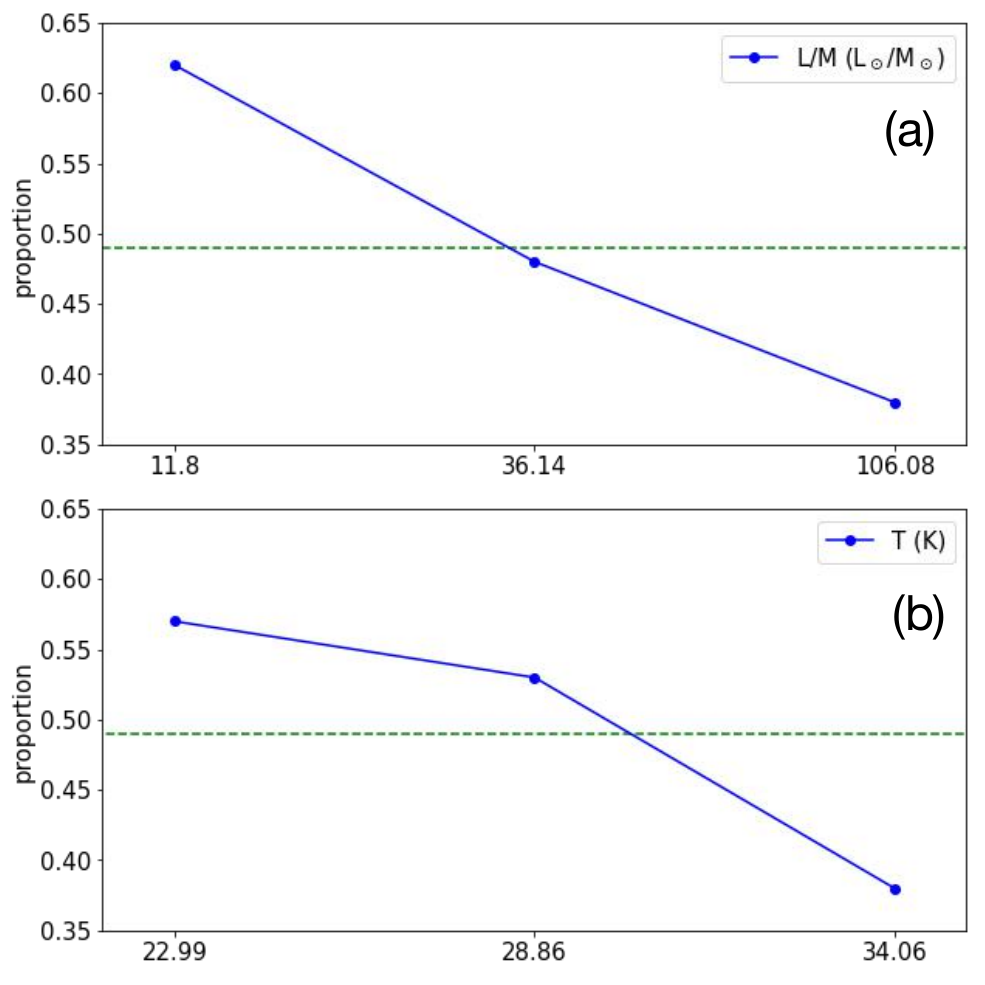}
\caption{Proportion of ATOM sources with hub-filament morphology in 3 groups divided by L/M and temperature T. (a) The bottom x axis represents the average value of L/M for all of sources in each group; (b) The bottom x axis is the average value of T. Dashed lines mark the proportion of hub-filament systems with good morphology for all of ATOMS sources.}
\label{group}
\end{figure}

As massive proto-stars evolve, the dust temperature ($T_d$) and luminosity-to-mass ratio ($L/M$) of their natal clumps will increase. Therefore, $T_d$ and $L/M$ are often used for the evolutionary classification of dense star-forming clumps \citep{Saraceno1996-309,Molinari2008-481,Liu2016-829, Stephens2016-824,Liuhongli2017,Urquhart2018-473,Molinari2019}. We divide ATOMS sources into three uniformly spaced bins in $T_d$ and $L/M$, respectively. We emphasize that $T_d$ and $L/M$ do not depend on distances in the sample (see Fig.~\ref{distance}b and Fig.~\ref{distance}c). Considering the small dynamic range in $T_d$ and $L/M$ of the data, we did not divide the sample into more bins, in order to avoid potential misleading results (see Appendix~\ref{appexA}). Then we calculate the proportion of sources with hub-filament morphology in each bin. As shown in Fig.~\ref{group}, the proportion of hub-filament systems decreases with increasing  $T_d$ and $L/M$, strongly indicating that hub-filament systems are gradually destroyed as proto-clusters evolve.

\citet{Liuh2021} identified all the dense cores in the ATOMS clumps and classified them based on their evolutionary stages (e.g., hot cores, UC H{\sc ii} regions). The evolutionary state of a clump can also be represented by the most evolved core within it. We find that nearly all the clumps (such as sources 25 and 63 in Fig.~\ref{hf25}) with their most evolved cores in hot molecular core phase or even earlier phases show very good hub-filament morphology. Some clumps (such as sources 44 and 71 in Fig.~\ref{hf25}) harboring UC H{\sc ii} regions also show robust hub-filament morphology, indicating that these UC H{\sc ii} regions are still confined by the dense gas in hub regions. As shown in Fig.~\ref{hi31}, expanding H{\sc ii} regions excited by formed proto-clusters will disperse the gas in the hub regions and destroy the hub-filament systems eventually. However, we also notice that those clumps associated with expanding UC H{\sc ii} regions or even more evolved H{\sc ii} regions can still sustain very good filamentary morphology, as shown in Fig.~\ref{hi31}.

To conclude, stellar feedback from H{\sc ii} regions gradually destroys the hub-filament systems as proto-clusters evolve.

\subsection{Velocity gradients along the longest filaments}\label{velocity}

\begin{figure*}
  \centering
  \includegraphics[width=0.95\textwidth]{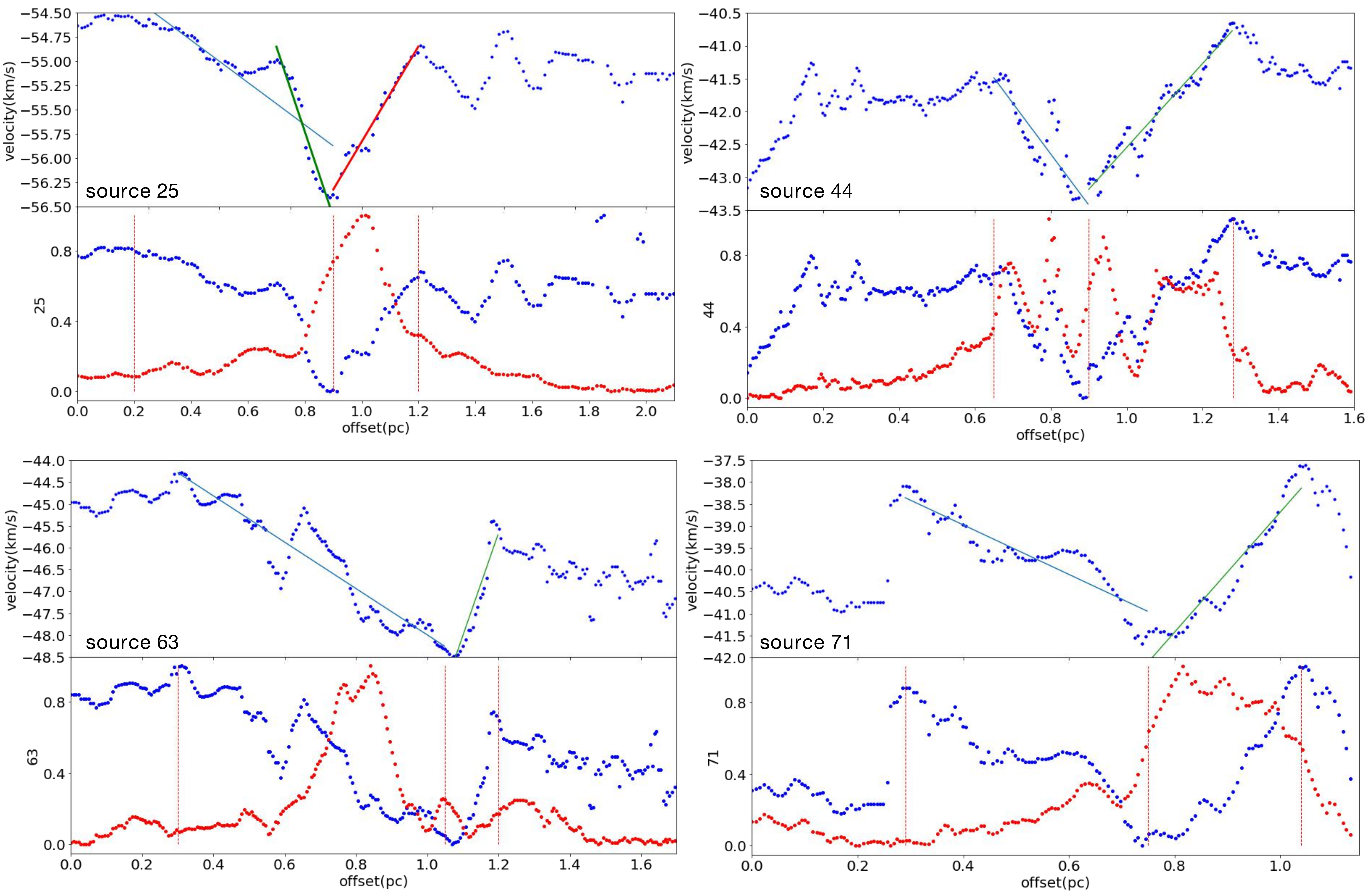}
\caption{The distributions of velocity and intensity along the longest filament in four exemplar sources. Upper panel for each source: Velocity gradients are fitted in the ranges defined by the red vertical dashed lines, and straight lines show the linear fitting results; Lower panel for each source: Blue and red dotted lines show the normalized velocity and intensity, respectively. The plots for all ATOMS sources are shown in the supplementary material.}
\label{wave}
\end{figure*}

Fig.~\ref{wave} shows the intensity-weighted velocity (Moment 1) and integrated intensity (Moment 0) of H$^{13}$CO$^{+}$ J=1-0 line emission along the longest filaments in four exemplar sources. Velocity and density fluctuations along these filaments are seen, which are likely caused by dense structures embedded in these filaments.  The density and velocity fluctuations along filaments may indicate oscillatory gas flows coupled to regularly spaced density enhancements that probably form via gravitational instabilities \citep{Henshaw2020}. We will do a more detailed analysis of this phenomena in a forthcoming paper.

On the other hand, one can see clear velocity gradients along the longest filaments. We firstly estimate two velocity gradients between velocity peaks and valleys at the two sides of the strongest intensity peaks of H$^{13}$CO$^+$ emission (i.e., the center of the gravity potential well), as marked by the vertical dashed lines in Fig.~\ref{wave}, and ignored local velocity fluctuation. We also derive additional velocity gradients over a smaller distances around the strongest intensity peaks of H$^{13}$CO$^+$ emission for some very long filaments (such as source 25 in the lower panels of Fig.~\ref{wave}). We note that the strongest intensity peaks of H$^{13}$CO$^+$ emission coincide with the brightest 3 mm cores or hub regions in those filament-hub systems. 

\begin{figure}
\centering
\includegraphics[scale=0.8]{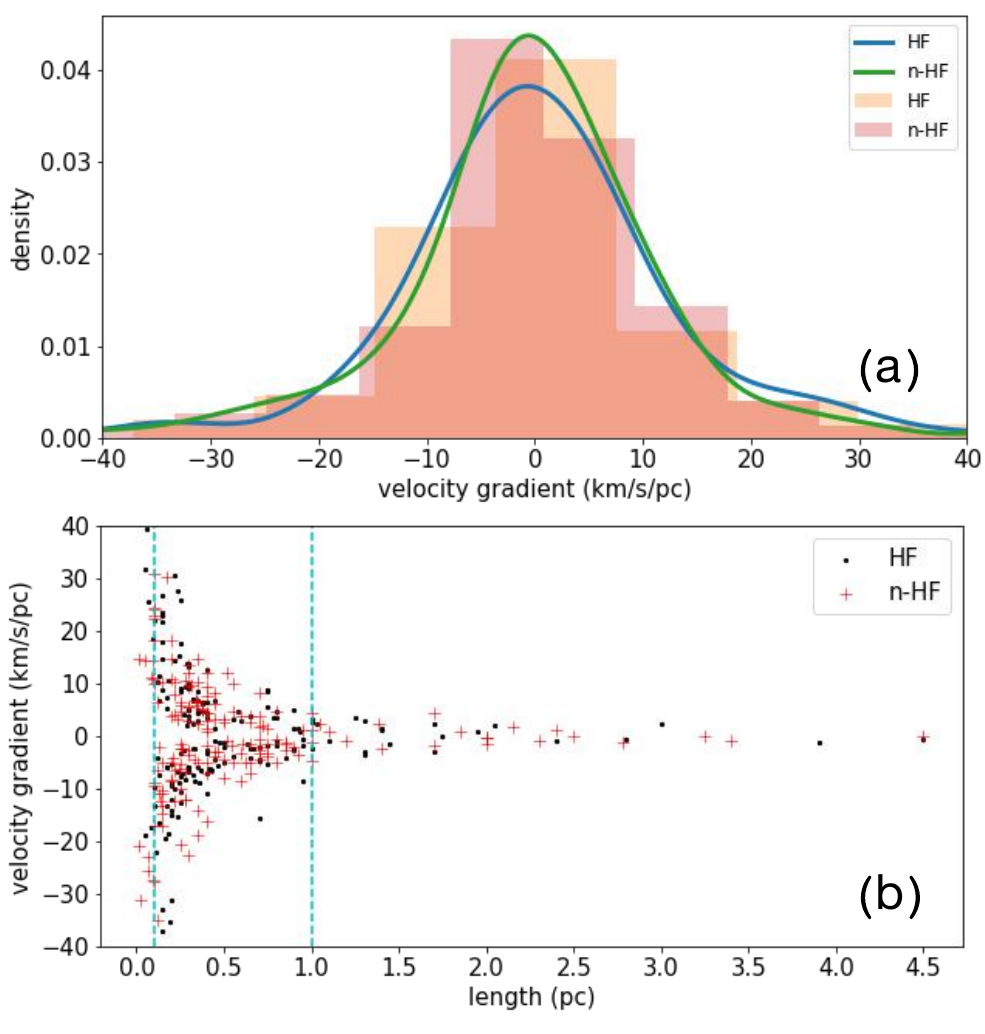}
\caption{(a) The distribution of velocity gradients along the longest filaments for all of ATOMS sources; (b) Velocity gradient versus the length over which the gradient has been estimated. Black dots and red plus signs represent the velocity gradients of hub-filament (HF) sources and non-hub-filament (n-HF) sources. Two cyan dashed lines mark the positions of 0.1pc and 1pc.}
\label{slope}
\end{figure}

In this work, we mainly focus on the longest filaments, which are the skeletons of the clumps. We did not study the other shorter filaments because not all shorter filaments are connected to the hubs. The shorter filaments will be investigated in future works. However, since we have a large sample, the total number of filaments in our analysis is still considerable from a statistical point of view. Fig.~\ref{slope}(a) shows the distribution of velocity gradients for all sources. The mean value and median value of velocity gradients are 8.71 km s$^{-1}$pc$^{-1}$ and 5.54 km s$^{-1}$pc$^{-1}$, respectively.  Statistically we can see approximately symmetric positive and negative velocity gradients in the distribution.

\begin{figure}
\centering
\includegraphics[scale=0.25]{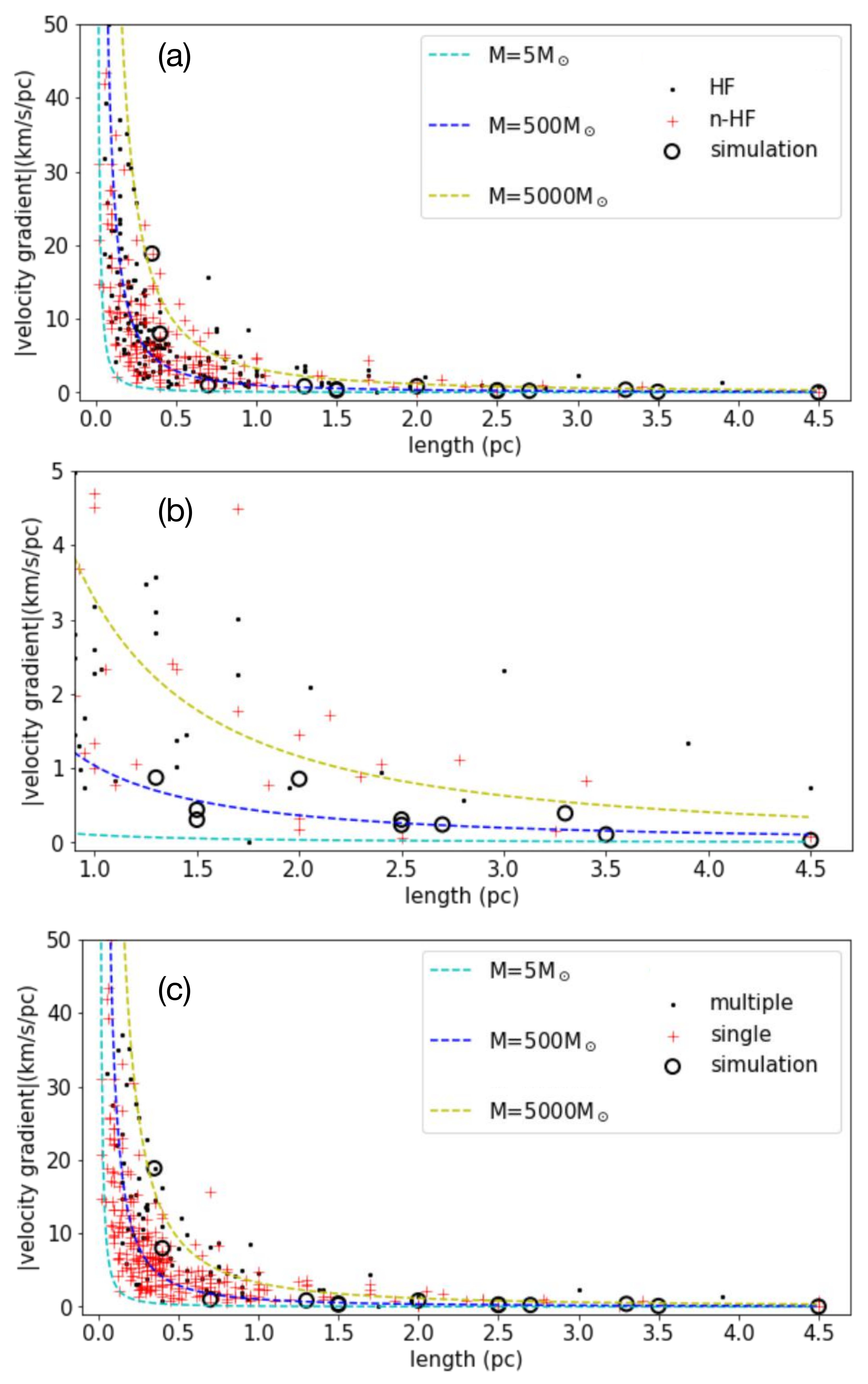}
\caption{(a) Velocity gradient versus the length over which the gradient has been estimated for all sources. The color lines show free-fall velocity gradients for comparison.  For the free-fall model, cyan, blue and yellow lines denote masses of 5M$_\odot$, 500M$_\odot$ and 5000M$_\odot$, respectively. Black circles represent the velocity gradients fitted by simulation data, as shown in Fig.\ref{simulation1}; (b) A blow-up of the region with lengths greater than 1 pc in panel a; (c) Same as in panel a, but sources showing singly-peaked profiles (red crosses) and multiple-peaked profiles (black dots) in H$^{13}$CO$^+$ line emission in hub regions are plotted separately. }
\label{free}
\end{figure}

Fig.~\ref{slope}(b) and Fig.~\ref{free} show velocity gradients as a function of the filament lengths over which the gradients have been estimated. There are no significant difference in velocity gradients among filaments with or without hubs. This may indicate that the physical drivers that generate the kinematics, such as filamentary accretion at different scales, are the same for filaments in various systems.
We find that velocity gradients are very small at scales larger than 1~pc relative to small scales ($<$1~pc), probably hinting for the existence of inertial inflow at large scales driven by turbulence \citep{Padoan2020}, or by large-scale gravitational collapse \citep{Gomez2014, Vazquez2019}, which compresses and pressurizes the filament, analogously to the compression of the innermost parts of a core by the infall of the envelope \citep{Gomez2021}. In the case of the filament, this pressure, together with the gravitational pull from the hub, may trigger the longitudinal flow. However, this can also originate from the natural attenuation of the gravitational pull from the hub at large distances from it along the filament, or by the combined gravity of the hub and the interior parts of the filament (cf.\ Sec.\ \ref{accretion}). This is evidenced by Fig.\ref{free}(b), which shows that the variation of velocity gradients on large scales more or less follow the trend at small scales. 

Below $\sim$1~pc, velocity gradients dramatically increase as filament lengths decrease, indicating that gravity dominates gas inflow at such small scales where gas is being accumulated in dense cores or hub regions. The large velocity gradients, however, could be overestimated due to the complicated line emission profiles in hub regions. Although the H$^{13}$CO$^+$ J=1-0 spectral lines over the mapping areas of the majority (75\%) of sources in the sample are singly peaked, we noticed asymmetric profiles or even double-peak profiles in lines toward the densest regions of some sources. These complicated line profiles may lead to high velocity gradients estimated from moment 1 maps near hub regions. Therefore, one should be very cautious about these exceptionally high velocity gradients (in several tens of km~s$^{-1}$). We note that this issue cannot be dealt with by simply fitting the spectra with multiple velocity components because they are more likely caused by the complicated gas motions in the hub regions rather than clearly separated velocity-coherent sub-structures. In Fig.~\ref{free}c, we separate the sources that show multiple-peaks in H$^{13}$CO$^+$ emission lines from those sources showing singly-peaked line profiles. From this plot, we found that the effect caused by the complicated line profiles in hub regions is not a severe problem in statistics, and the trend in the relation between velocity gradients and filament lengths is not changed. 

In a recent review work by \cite{Hacar2022}, they also found similar trend of the variation of velocity gradients as a function of filament lengths as we witnessed here (Fig.\ref{free}). However, they did not include such massive hub-filament systems from high-resolution interferometric observations in their study.

\section{Discussion}
\label{sec:discussion}

\subsection{From gas inflow to infall along filaments}\label{accretion}

In observations, velocity gradients along filaments are often interpreted as evidence for gas inflow along filaments \citep{Kirk2013,Liu2016,Yuan2018,Williams2018,Chen2019,ChenM2020,Pillai2020}. 

As mentioned above, the longitudinal flow along the filaments may be due to either a pressure-driven flow or to self-gravity, or a combination thereof. However, each of these possibilities in turn splits in two possible sub-cases. In the case of the inertial flow, it can be driven by the large-scale turbulent ram pressure, as suggested by \citet{Padoan2020}, or to compression by a cloud-scale gravitational contraction, as suggested in the GHC model of \citet{Vazquez2019}. On the other hand, if due to self-gravity, it can be dominated by the hub, or by the filament, or both. We consider these possibilities in the Appendix \ref{theory}, and find that longitudinal velocity profile can be naturally explained by the filament’s self-gravity.

\subsubsection{Longitudinal flow along filaments in simulations.}

This idea of longitudinal flows along filaments can also be tested in simulations.

\begin{figure*}
  \centering
  \includegraphics[width=0.9\textwidth]{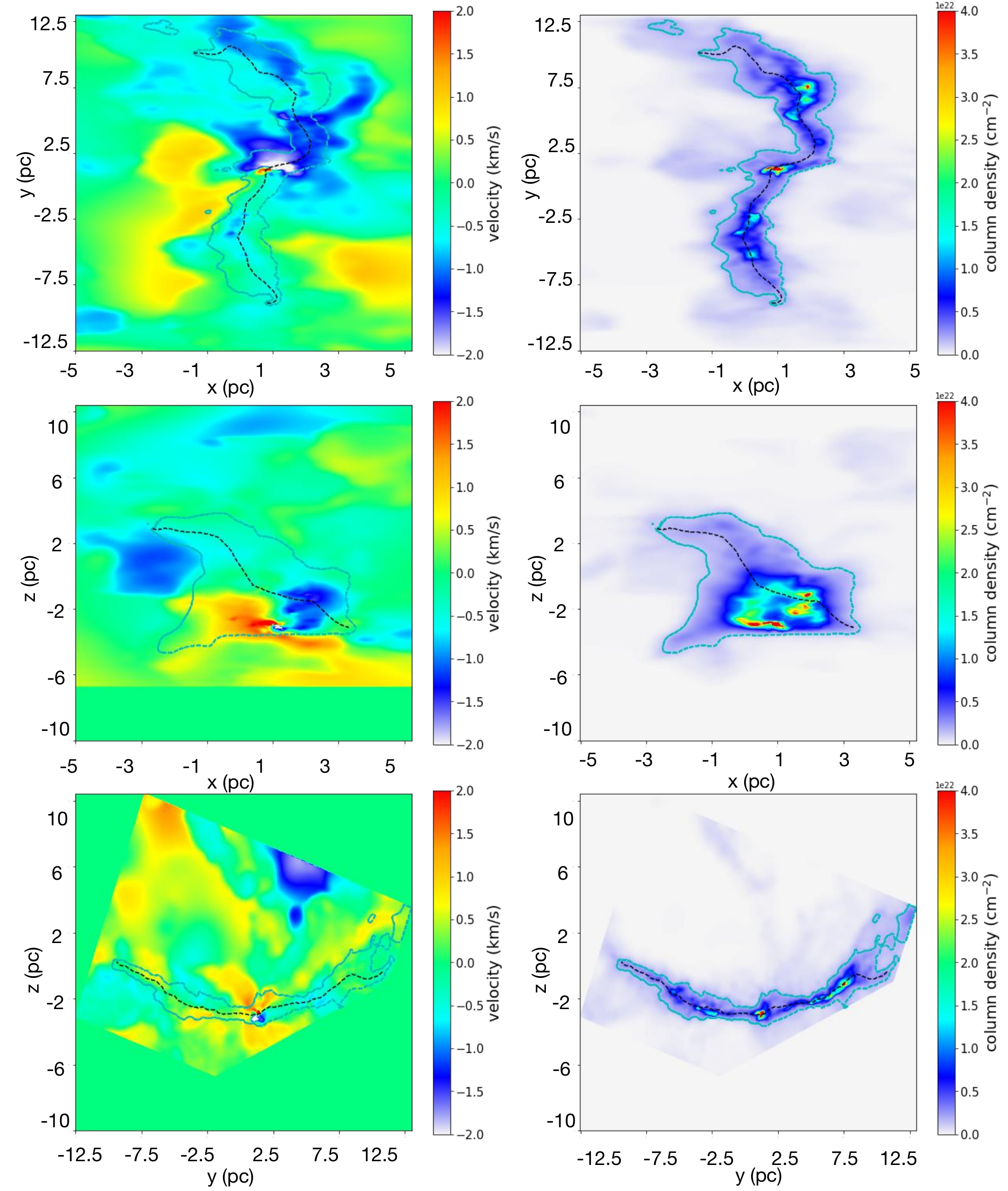}
\caption{The distributions of velocity and intensity in three projection planes of three dimensional SPH simulation in \citet{Gomez2014}. Blue dashed lines show $\sim$4$\sigma$ column density contours, black dashed line represent the longest filament identified by FILFINDER according to the column density map.}
\label{simulation2}
\end{figure*}

\citet{Gomez2014} performed SPH simulations of the formation of a molecular cloud from a convergent flow of diffuse gas. Long filaments with lengths up to 15 pc are formed in the cloud as a consequence of anisotropic, gravitationally-driven contraction flow. Fig.~\ref{simulation2} presents the density and density-weighted velocity maps of one cloud in their simulation, viewed from three different angles. The cloud exhibits clear filamentary morphology as viewed along z-axis or x-axis. However, the filament morphology is not obviously seen in the x-z plane. This indicates that projection effect in observations cannot be ignored in interpreting filament properties. Since the longest filaments in our observations have large aspect ratios, these filaments in our studies do not seem to have extreme inclination angles.

\begin{figure*}
  \centering
  \includegraphics[width=0.9\textwidth]{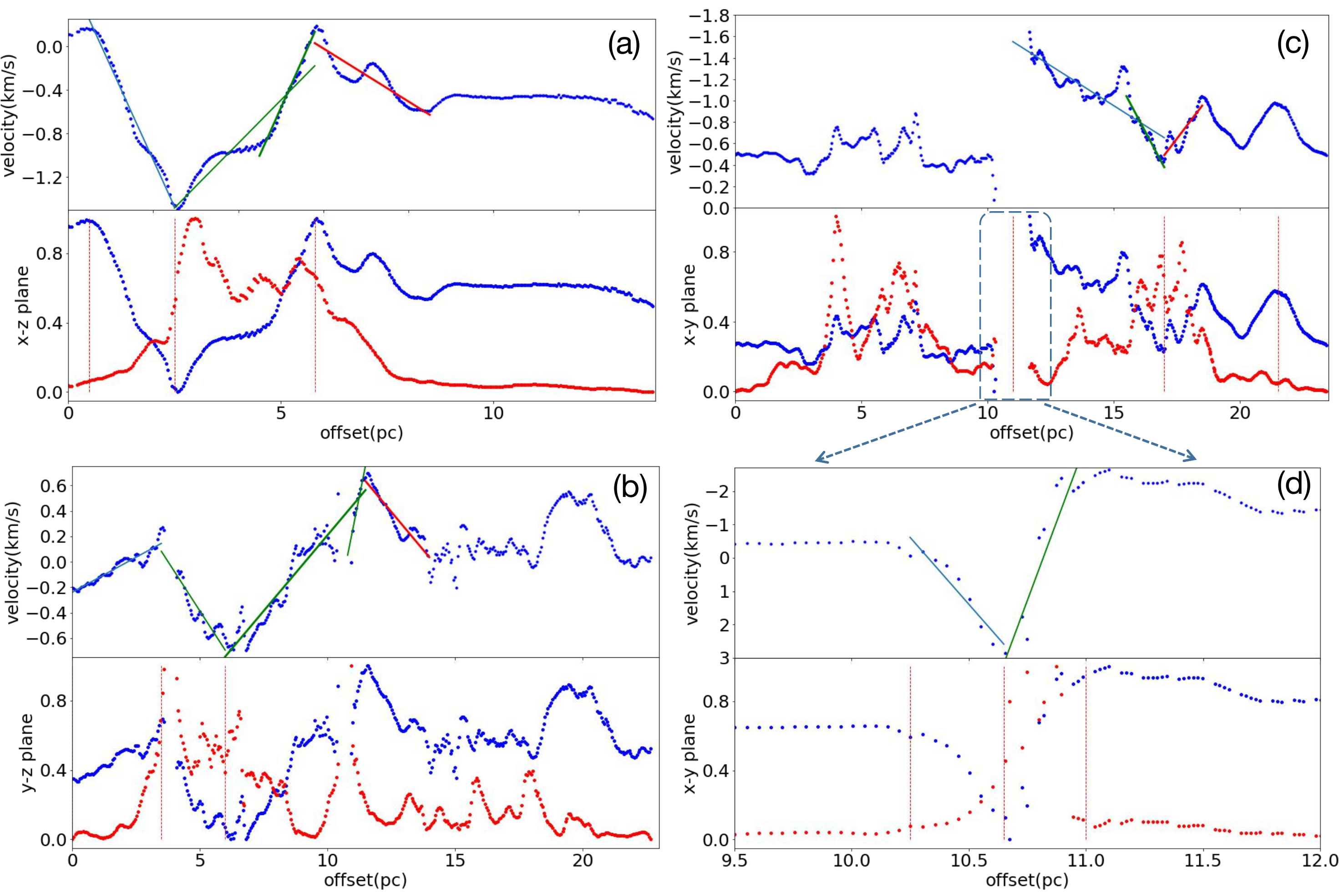}
\caption{The distributions of velocity and intensity along the longest filament marked by black dashed lines in Fig.\ref{simulation2}. Their features are the same with Fig.\ref{wave}. (d) shows the velocity shear region marked in (c) by blue dashed box.}
\label{simulation1}
\end{figure*}

\citet{Gomez2014} witnessed clear longitudinal inflow along filaments in their simulations. Fig.\ref{simulation1} presents the distributions of density and density-weighted velocity along the skeletons of filaments in their simulation. Density and velocity fluctuations are also seen in these simulated filaments. In addition, clear velocity gradients at both large-scale ($>$1 pc) and small scale ($<$1 pc) are also seen around density peaks. These patterns are remarkably similar to our observational results as seen in Fig.~\ref{wave}. We also derived the overall velocity gradients and local velocity gradients for this simulated filament. In general, the local velocity gradients around density peaks are much larger than the overall velocity gradients at large scales no matter how the filament orients with different inclination angles. The velocity gradients in simulated data are more or less consistent with those in our observations as seen in Fig.~\ref{free}. This consistency strongly suggests that the velocity gradients along the longest filaments in ATOMS sources are likely caused by longitudinal inflow.

In addition, large-scale velocity gradients perpendicular to the main filament exist across the whole cloud in simulations as seen in the velocity maps of Fig.~\ref{simulation2}. This indicates that the filament itself is formed due to a convergent flow of diffuse gas in its environment. Velocity gradients perpendicular to filaments are also seen in ATOMS sources as shown in the moment 1 maps of Fig.~\ref{hf25}. In a thorough case study of an ATOMS source, G286.21+0.17, \citet{Zhou2021} revealed prominent velocity gradients perpendicular to the major axes of its main filaments, and argued that the filaments are formed due to large-scale compression flows, possibly driven by nearby H{\sc ii} regions and/or cloud-scale gravitational contraction.

\subsubsection{Gas infall near hub regions or dense cores}

Considering projection effects, the filament lengths would approximate the distances to the centers of mass concentrations (hubs or cores). Therefore, the velocity gradient measured along a filament can be treated as velocity gradient at a contain distance from the center of gravity potential well.  As shown in Fig.~\ref{slope}(b) and Fig.~\ref{free},  velocity gradients dramatically increase as filament lengths decrease at scales smaller than 1~pc. This indicates that the hub's or core's gravity dominates gas flow at such small scales. The observed velocity gradients can be compared with that of free-fall. The free-fall velocity gradient is:
\begin{equation}
\nabla V_{free}= -\frac{d}{dR}\sqrt{\frac{2GM}{R}}
=\sqrt{\frac{GM}{2R^3}}.
\label{v-free}
\end{equation}
where $M$ is the mass of the gas concentrations (hubs or cores) and R is the distance to their gravity potential centers. Since the sizes of hubs or cores are far smaller than the filament lengths, here we treat the hubs or cores as point objects.
As seen from Fig.~\ref{free}, the observed velocity gradients roughly follow the free-fall models at small-scales, indicating the existence of gas infall governed by the gravity of the hub, or of dense cores located along filaments, in their immediate surroundings. We note that free-fall model may not be realistic, and we also did not consider the masses of filaments themselves. However, this simple comparison indicates that gravity could dominate gas infall at small scales.

\begin{figure}
\centering
\includegraphics[scale=0.9]{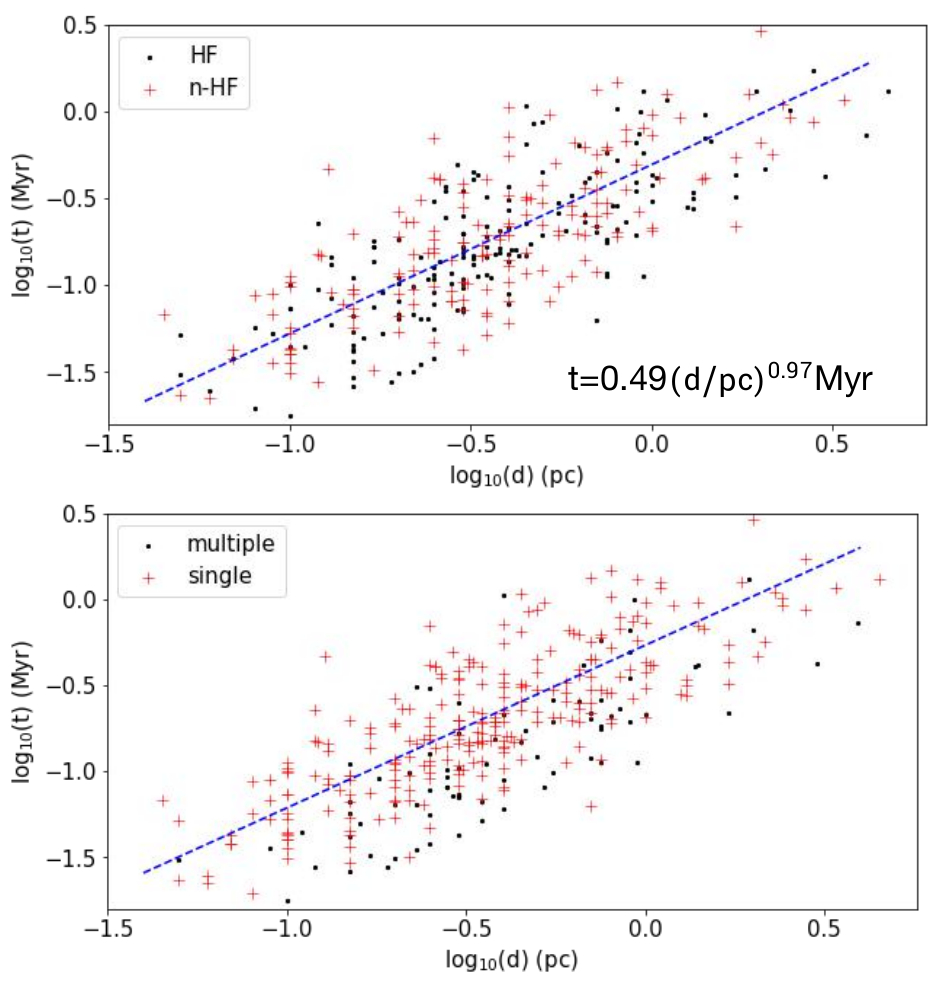}
\caption{Gas accretion timescale $t_{inflow}$ versus filament length $d_{fila}$ over which the velocity gradient is estimated. (a) Hub-filament systems (black dots) and non Hub-filament systems (red crosses); (b) sources (red crosses) showing singly-peaked profiles of H$^{13}$CO$^+$ line emission in hub regions and sources (black dots) showing multiple-peaked profiles. }
\label{time}
\end{figure}

\subsubsection{The timescale of filamentary accretion}

Below we estimate the gas accretion timescale of the longest filaments in the ATOMS proto-clusters. Assuming that the velocity gradient along filament is caused by gas inflow, the mass inflow rate ($\dot{M}_{\parallel}$) along the filament can be estimated by:
\begin{equation}
\dot{M}_{\parallel} = \frac{\nabla V_{\parallel,obs} M}{\tan(\alpha)}
\end{equation}
where $\nabla V_{\parallel,obs}$ is the velocity gradient along the filament, $M$ is the filament mass and $\alpha$ is the inclination angle of the filament relative to
the plane of the sky \citep{Kirk2013}. 
The gas accretion time $t_{inflow}$ scale is then given by:
\begin{equation}
t_{inflow}=\frac{M}{\dot{M}_{\parallel}} = \frac{\tan(\alpha)}{\nabla V_{\parallel,obs}}
\end{equation}
In calculating $t_{inflow}$, we simply take $\alpha$ equaling to 45$^{\circ}$ because the longest filaments do not seem to have extreme inclination angles. Therefore, the correction factor for $t_{inflow}$  considering projection effect should be only a factor of a few. The mean value and median value for the gas accretion time are 0.25 Myr and 0.16 Myr, respectively.

We plot $t_{inflow}$ as a function of filament length $d_{fila}$ (over which the velocity gradient is estimated) in Fig.~\ref{time}. From this figure, we find that $t_{inflow}$ is tightly correlated with $d_{fila}$ in spite of the large data scatter. 
The correlation can be well fitted by a nearly linear relation: 
\begin{equation}
t_{inflow}=0.49 ({d_{fila}/{\rm pc}})^{0.97} {\rm Myr},
\label{t4}
\end{equation}
The slope 0.97 is very close to 1 with a correlation coefficient r=0.75.

\subsection{Comparison with theoretical models}\label{models}

There are several popular models for explaining high-mass formation. However, none of them has been fully proven. 

The turbulent-core model \citep{McKee2003,Krumholz2007} assumes that the formation of a massive star starts from the collapse of a massive prestellar core, which is the gas reservoir for the final stellar mass. However, the prevalent feature of hub-filament systems and potential large-scale converging gas inflows along filaments at scales from $\sim$0.1 pc to several pc discovered in this work support the idea that the gas reservoirs for high-mass star formation may not be pre-existing massive turbulent cores. Rather, the hubs and dense cores themselves may accumulate most of their masses through large-scale filamentary accretion. 

The "clump-fed" models, such as competitive-accretion model \citep{Bonnell1997,Bonnell2001}, inertial-inflow model \citep{Padoan2020}, and global hierarchical collapse model \citep{Vazquez2009,Ballesteros2011,Hartmann2012,Vazquez2017,Vazquez2019}, assume that high-mass stars are born with low stellar masses but grow to much larger final stellar masses by accumulating mass from large-scale gas reservoirs beyond their natal dense cores. In these models, large--scale converging flows or global collapse of clumps are required to continuously feed mass into the dense cores. However, the method of accretion or accumulating material in these models are different. 

The competitive-accretion model only accounts for mass accretion due to the gravity of the growing proto-stars (Bondi-Hoyle accretion), neglecting the preexisting inflow at the larger scales as we observed here. 

In the inertial-inflow model, the prestellar cores that evolve into massive stars have a broad mass distribution but can accrete gas from parsec scales through large--scale converging flows \citep{Padoan2020}. This model predicts that the parsec-scale region around a prestellar core is turbulent and gravitationally unbound \citep{Padoan2020}. However, in observations, most Galactic parsec-scale massive clumps seem to be gravitationally bound no matter how evolved they are \citep{Liu2016-829,Urquhart2018-473}. The inertial-inflow model also predicts that the net inflow velocity in inflow region is generally much smaller than the turbulent velocity and is not dominated by gravity \citep{Padoan2020}. This is contrary to our results as shown in Fig.\ref{slope}(b), from which one can see that gravity may start to dominate gas infall at scales smaller than $\sim$1 pc.

The global hierarchical collapse (GHC) model advocates a picture of molecular clouds in a state of hierarchical and chaotic gravitational collapse (multi-scale infall motions), in which local centers of collapse develop throughout the cloud while the cloud itself is also contracting \citep{Vazquez2019}. This model predicts anisotropic gravitational contraction with longitudinal flow along filaments at all scales. In our work, we found that velocity gradients along filaments are small at scales larger than $\sim$1 pc, indicating that gravity may not dominate gas flow at such large parsec scale. However, both GHC model and inertial-inflow predict small velocity gradients at large scales, and they are not distinguishable in this aspect. 

In conclusion, the prevalent hub-filament systems found in proto-clusters favors the pictures advocated by either global hierarchical collapse or inertial-inflow scenarios, which emphasize longitudinal flow along filaments. However, here we argue that gas infall at scales smaller than $\sim$1 parcsec is likely dominated by the gravity of the hub, while velocity gradients are very small at scales larger than $\sim$1~pc, probably suggesting the dominance of either pressure-driven inflow, which may be driven either by large-scale turbulent motions (inertial inflow) or large-, cloud-scale collapse, or else the self-gravity of the filament. Similar results can also be seen in \citet{Liu2021arXiv211102231L,Liu2022b}.

\begin{figure}
\centering
\includegraphics[scale=0.35]{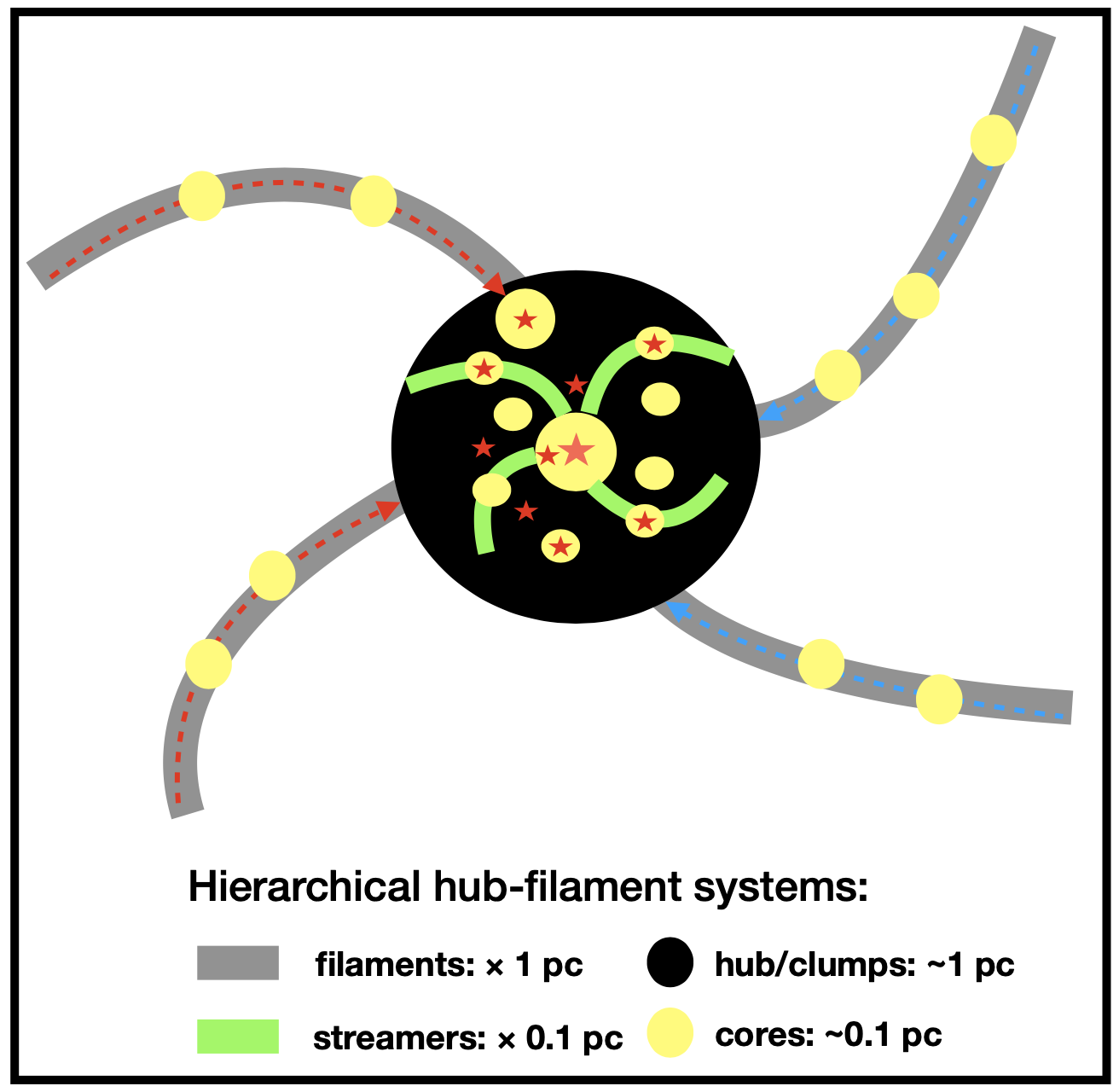}
\caption{Schematic diagram of self-similar hierarchical hub-filament systems at different scales.}
\label{s-similar}
\end{figure}

As discussed in Sec.\ref{Ubiquitous}, we find that hub-filament structures can exist at various scales from 0.1 parsec to several parsec in very different Galactic environments as described in Fig.\ref{s-similar}. Interestingly, below the 0.1 parcsec scale, slender structures similar to filaments, such as spiral arms, have also been detected in the surroundings of high-mass protostars \citep{Liu2015-804,Maud2017-467,Izquierdo2018-478,Chen2020-4,Sanhueza2021-915}. Therefore, self-similar hub-filament systems and filamentary accretion seem to exist at all scales (from several thousands au to several parsec) in high-mass star forming regions . This paradigm of hierarchical hub-filament like systems should be considered in any promising model for high-mass star formation.

\subsection{Caveats in this work}\label{caveats}

In this work, we identified filaments using H$^{13}$CO$^+$ J=1-0 data with the FILFINDER algorithm. However, the way that individual filaments have been defined is completely dependent on the algorithm that has been used. In FILFINDER algorithm, there are two key parameters, skel-thresh and branch-thresh, which affect the number of identified skeletons and branches very much. Although we carefully adjusted these parameters for individual sources, we cannot guarantee that the filament networks identified in our study ideally unveil the underlying hierarchical structures of the clumps. However, the identification of the longest filament skeletons is not greatly affected by choosing different parameters. Therefore, we mainly investigate the gas kinematics of the longest filaments in this work. 

In addition, although H$^{13}$CO$^+$ J=1-0 is a good tracer of dense gas, it may not be able to trace the highest density gas due to its relatively low critical density ($\sim$2$\times10^4$ cm$^{-3}$). Future studies with various gas tracers that have different excitation conditions could be helpful to fully reveal the gas distribution within these massive clumps.

The other caveat in this work is about the classification of hub-filament systems. The exact number of hub-filament systems in the sample is very dependent on the definition. The classification of hub-filament systems by eye is also not completely reliable. However, from a statistical view with such a large sample, our analysis should not be greatly affected by these issues. 

In this work, we interpret velocity gradients along filaments as evidences for gas inflow or gas infall. Basically, this may be true because most clumps in the sample are gravitationally bound with small virial parameters \citep{Liu2016-829,Liu2020b}. However, we have ignored the effects of stellar feedback such as outflows, radiation and/or stellar winds in H{\sc ii} regions, which may cause outward motions along or across filaments. The complex interplay between gas inflow and stellar feedback will be investigated thoroughly in future works.

\section{Summary}
\label{sec:summary}
We have studied the physical properties and evolution of hub--filament systems in a large sample of proto-clusters that were observed in the ATOMS survey. The main results of this work are as follows:

(1) We use the Moment 0 maps (integrated intensity maps) of the H$^{13}$CO$^{+}$ J=1-0 emission and the FILFINDER algorithm to identify filaments in ATOMS sources. We find that filaments are nearly ubiquitou in proto-clusters. With our strict definition, 49$\%$ of sources are classified as hub-filament systems. 

(2) We find that hub-filament structures can exist not only in small--scale ($\sim$0.1 pc) dense cores but also in large--scale clumps/clouds ($\sim$1-10 pc), suggesting that self-similar hub-filament systems at various scales are crucial for star and stellar cluster formation in various Galactic environments. 

(3) We find that the proportion of hub--filament systems decreases as $T_d$ and $L/M$ increases, indicating that stellar feedback from H{\sc ii} regions gradually destroys the hub-filament systems as proto-clusters evolve. Hub-filament systems in clumps containing H{\sc ii} regions may have been destroyed by stellar feedback as H{\sc ii} regions expand. Therefore, we argue hub-filament systems are crucially important in the formation and evolution of massive proto-clusters. 

(4)  The longest filaments in ATOMS sources show clear velocity gradients. The approximately symmetric distribution of positive and negative velocity gradients strongly indicates the existence of converging gas inflows along filaments. We also find that velocity gradients are very small at scales larger than $\sim$1~pc, probably suggesting the dominance of  pressure--driven inertial inflow, which can originate either from large-scale turbulence or from cloud-scale gravitational contraction. Below $\sim$1~pc, velocity gradients dramatically increase as filament lengths decrease, indicating that the hub's or core's gravity dominates gas infall at such small scales. Assuming that the velocity gradients along filaments are caused by gas inflow, we find that gas inflow timescale $t_{inflow}$ is linearly correlated with filament length $d_{fila}$. Our observations indicate that high-mass stars in proto-clusters may accumulate most of their mass through longitudinal inflow along filaments.

(5) We argue that any promising models for high-mass star formation should include self-similar hub-filament systems and filamentary accretion at all scales (from several thousand au to several parsec).

\section*{Acknowledgements}
This paper makes use of the following ALMA data: ADS/JAO.ALMA$\sharp$2019.1.00685.S. ALMA is a partnership of ESO (representing its member states), NSF (USA) and NINS (Japan), together with NRC (Canada), MOST and ASIAA (Taiwan), and KASI (Republic of Korea), in cooperation with the Republic of Chile. The Joint ALMA Observatory is operated by ESO, AUI/NRAO and NAOJ.

Tie Liu acknowledges the supports by National Natural Science Foundation of China (NSFC) through grants No.12073061 and No.12122307, the international partnership program of Chinese Academy of Sciences through grant No.114231KYSB20200009, Shanghai Pujiang Program 20PJ1415500 and the science research grants from the
China Manned Space Project with no. CMS-CSST-2021-B06.

NJE thanks the Department of Astronomy at the University of Texas at Austin for ongoing research support.

D. Li is supported by the National Natural Science Foundation of China grant No. 11988101

C. W. L. is supported by the Basic Science Research Program through the National Research Foundation of Korea (NRF) funded by the Ministry of Education, Science and Technology (NRF-2019R1A2C1010851).

L.B. and G.G. gratefully acknowledge support by the ANID BASAL projects ACE210002 and FB210003.

E.V.-S. acknowledges financial support from CONACYT grant 255295.

This research was carried out in part at the Jet Propulsion Laboratory, which is operated by the California Institute of Technology under a contract with the National Aeronautics and Space Administration (80NM0018D0004).

S.-L. Qin is supported by the National Natural Science Foundation of China (grant No. 12033005).

H.-L. Liu is supported by National Natural Science Foundation of China (NSFC) through the grant No.12103045.

JHH thanks the National Natural Science Foundation of China under grant Nos. 11873086 and U1631237.
This work is sponsored (in part) by the Chinese Academy of Sciences (CAS), through a grant to the CAS South America Center for Astronomy (CASSACA) in Santiago, Chile.

G.C.G. acknowledges support by UNAM-PAPIIT IN103822 grant.

Zhiyuan Ren is supported by NSFC E013430201, 11988101, 11725313, 11403041, 11373038, 11373045
and U1931117.

S. Zhang acknowledges the support of China Postdoctoral Science Foundation through grant No. 2021M700248.

K.T. was supported by JSPS KAKENHI (Grant Number 20H05645). 

TB acknowledge the support from S. N. Bose National Centre for Basic Sciences under the Department of Science and Technology (DST), Govt. of India.

YZ wishes to thank the National Science Foundation of China (NSFC, Grant No. 11973099) and the science research grants from the China Manned Space Project (NO. CMS-CSST-2021-A09 and CMS-CSST-2021-A10) for financial supports.

JG thanks the support from the Chinese Academy of Sciences (CAS) through a Postdoctoral Fellowship administered by the CAS South America Center for Astronomy (CASSACA) in Santiago, Chile.

C.E. acknowledges the financial support from grant RJF/2020/000071 as a part of Ramanujan Fellowship awarded by Science and Engineering Research Board (SERB), Department of Science and Technology (DST), Govt. of India.

\section{Data availability}The data underlying this article are available in the article and in ALMA archive.

\bibliography{HF}{}
\bibliographystyle{aasjournal}

\appendix

\section{Distances and evolution of Hub-filament systems}\label{appexA}

\begin{figure}
\centering
\includegraphics[scale=0.95]{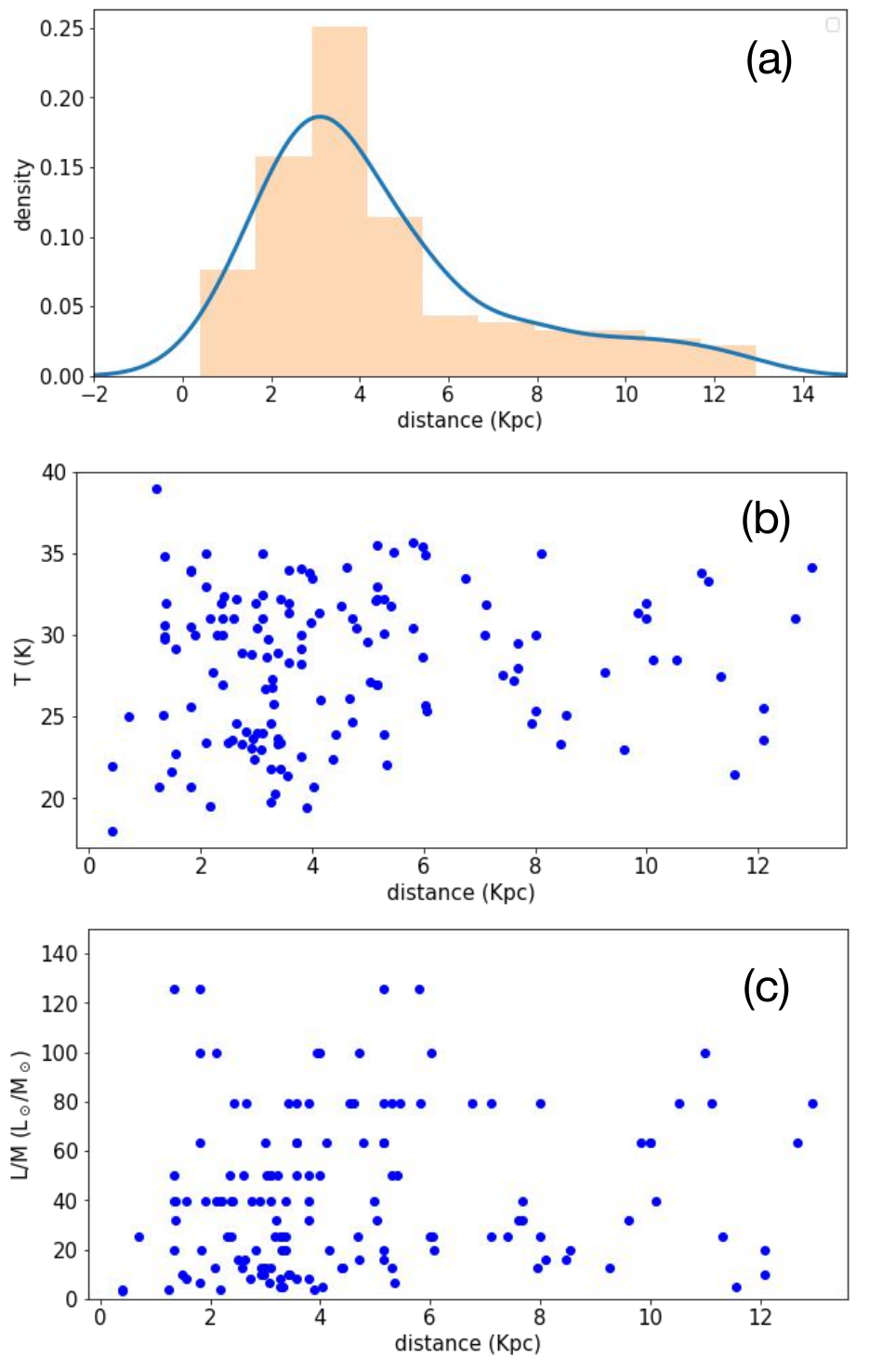}
\caption{(a) Distance distribution of ATOMS sources; (b) Dust temperature of ATOMS sources versus distance; (c) L/M of ATOMS sources versus distance.}
\label{distance}
\end{figure}

Panel (a) in Fig.~\ref{distance} shows the distribution of distances for the ATOMS sample. Panel (b) in Fig.~\ref{distance} presents dust temperature ($T_d$) versus distance for all ATOMS sources, and panel (c) shows the relation between luminosity-to-mass ratios ($L/M$) and distances. From these two panels, one can see that $T_d$ and L/M do not depend on distances in the ATOMS sample.

\begin{figure}
\centering
\includegraphics[scale=0.9]{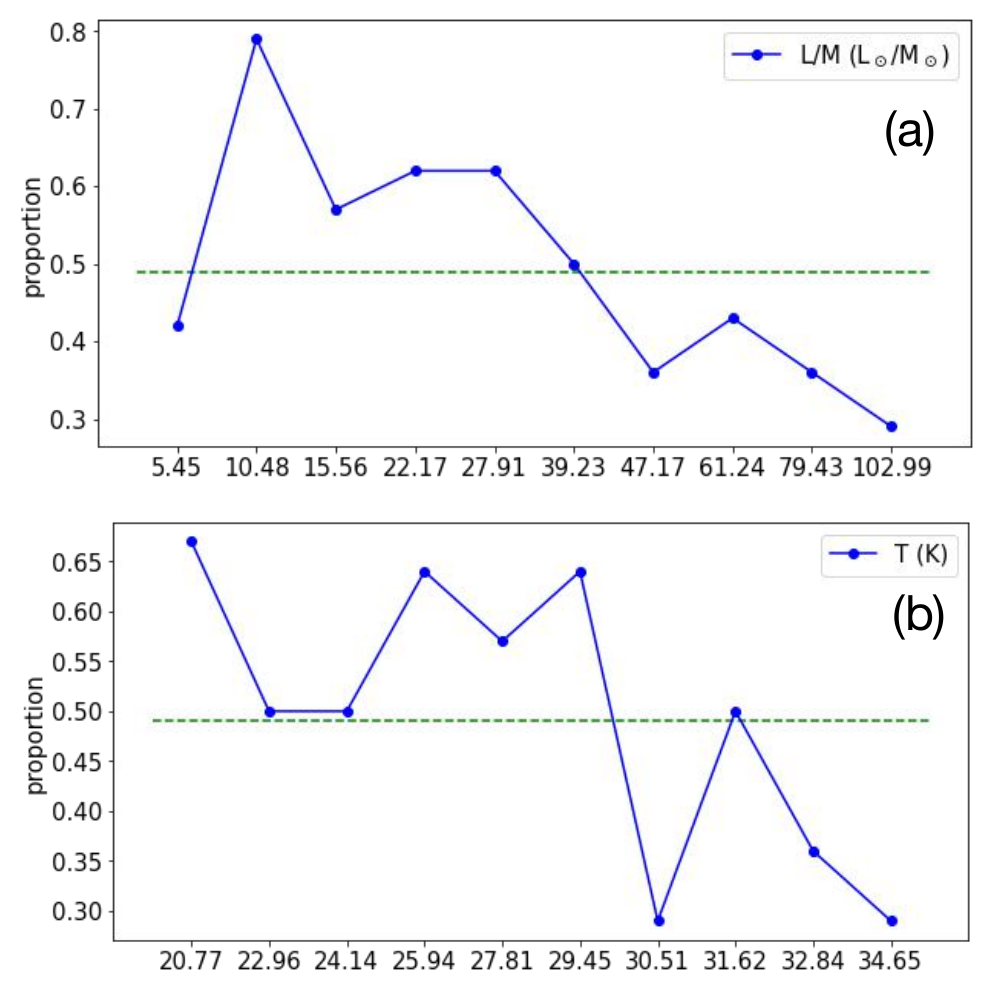}
\caption{Proportion of ATOM sources with hub-filament morphology divided by L/M and temperature T in 10 bins. (a) The bottom x axis represents the average value of L/M for all of sources in each bin; (b) The bottom x axis is the average value of T. Dashed lines mark the proportion of hub-filament systems with good morphology for all of ATOMS sources.}
\label{morebins}
\end{figure}

In Fig.~\ref{morebins}, we divide ATOMS sources into ten uniformly spaced bins in $T_d$ and $L/M$, respectively. As shown this figure, the proportion of hub-filament systems decreases with increasing  $T_d$ and $L/M$ in general. We noticed that the sources with smallest $L/M$ have less hun-filament systems, which may indicate that hub-filament systems have not been fully developed in the youngest clumps. However, these results should be taken caution because each bin only contains about 14 sources and the statistics with such a small number of sources in each bin may not be significant. Studies with a larger sample of sources with a larger dynamic range in $T_d$ and $L/M$ would help further constrain the evolution of hub-filament systems.

\section{Longitudinal velocity profile in theory}\label{theory}

\subsection{The longitudinal velocity profile for a constant accretion rate per unit length onto the filament}

Let us consider the pressure-driven case first. Without making any assumptions on whether the pressure is due to turbulence or gravitational infall from the cloud scale, let us consider the case of a filament of linear mass density $\lambda$, which hosts a longitudinal flow with velocity $v(z)$, where $z$ is the coordinate along the filament. In addition, we assume that the filament is accreting radially from the cloud at a uniform rate per unit length $\dot \lambda_0$ throughout its length, so that each cylindrical segment of thickness $\Delta z$ of the filament accretes mass radially at a rate
\begin{equation}
\dot M_{\rm rad} = \dot \lambda_0\, \Delta z.
\label{dotM_rad}
\end{equation}
However, as seen from the intensity (red dotted) lines in Fig.\ \ref{wave}, the linear mass density of the filament appears to be roughly constant to zeroth order along the filament. To achieve this constancy,  the longitudinal accretion rate $\lambda v$ must increase across $\Delta z$ by an amount 
\begin{equation}
\Delta \dot M_{\rm long} = \lambda \Delta v = \dot M_{\rm rad} = \dot \lambda_0\, \Delta z.
\label{delta_dotM_long}
\end{equation}
Since both $\lambda$ and $\dot \lambda_0$ are constants to zeroth order, we the find that, to this order, 
\begin{equation}
\Delta v \propto \Delta z,
\label{DvDz}
\end{equation}
which is consistent with the linear fits to the velocity seen in Fig.\ \ref{wave}.

\subsection{The longitudinal velocity profile due to the filament's self-gravity}

Let us consider an infinitely thin filament of uniform linear mass density $\lambda$ placed between $-L$ and $L$,
away from the gravitational force of the hub or early in the filament evolution, i.e., before the hub formation.
Consider also a gas parcel within the filament at $z$ and, for simplicity, assume that $z > 0$.
The gravitational force experienced by the gas parcel is given by,

\begin{align}
    F(z) &= \int_{-L}^{2z-L} \frac{G \lambda \dif z'}{(z'-z)^2} \nonumber \\
         &= \frac{-2 G \lambda z}{L^2-z^2},
    \label{eq:fil_force}
\end{align}

\noindent
(since the force due to the $[z,L]$ filament segment is balanced by the symmetrical $[2z-L,z]$ one.
See \citealt{Burkert2004} for a more detailed analysis).
The work $\Delta W$ exerted by this force onto the parcel will give it a kinetic energy $K$.
Assuming that the gas parcel starts at $z_0 \ll L$ at zero velocity, then $\Delta W = K$ implies,

\begin{equation}
    v^2(z) = 2 G \lambda (z_0^2 - z^2),
    \label{eq:fil_velocity}
\end{equation}

\noindent
and the velocity of the gas parcel will be approximately linear with $z$.

Notice that in this simple model, the velocity vector at either side of the filament center should point to $z=0$, so it should be interpreted as modeling the flow associated with individual collapse centers embedded within the filament.
In this sense, the velocity profile in eq. \eqref{eq:fil_velocity} calculated for a finite filament is applicable to segments surrounding individual collapses within the filament as a whole, highlighting the self-similar nature of the process.

\clearpage

\noindent
Author affiliations:\\
$^{1}$National Astronomical Observatories, Chinese Academy of Sciences, Beijing 100101, Peoples Republic of China \\
$^{2}$University of Chinese Academy of Sciences, Beijing 100049, Peoples Republic of China  \\
$^{3}$Shanghai Astronomical Observatory, Chinese Academy of Sciences, 80 Nandan Road, Shanghai 200030, Peoples Republic of China \\
$^{4}$Department of Astronomy, The University of Texas at Austin,
2515 Speedway, Stop C1400, Austin, Texas 78712-1205, USA\\
$^{5}$Departamento de Astronom{\i}a, Universidad de Chile, Camino el Observatorio 1515, Las Condes, Santiago, Chile \\
$^{6}$Jet Propulsion Laboratory, California Institute of Technology, 4800 Oak Grove Drive, Pasadena CA 91109, USA\\
$^{7}$Department of Physics, P.O.Box 64, FI-00014, University of Helsinki, Finland\\
$^{8}$Department of Astronomy, Yunnan University, Kunming, 650091, PR China\\
$^{9}$Instituto de Radioastronomía y Astrofísica, Universidad Nacional Autónoma de México, Antigua Carretera a Pátzcuaro \# 8701, Ex-Hda. San José de la Huerta, Morelia, Michoacán, México C.P. 58089\\
$^{10}$Yunnan Observatories, Chinese Academy of Sciences, 396 Yangfangwang, Guandu District, Kunming, 650216, P. R. China\\
$^{11}$Chinese Academy of Sciences South America Center for Astronomy, National Astronomical Observatories, CAS, Beijing 100101, China\\
$^{12}$Departamento de Astronom\'{i}a, Universidad de Chile, Casilla 36-D, Santiago, Chile\\
$^{13}$Departamento de Astronom\'{i}a, Universidad de Concepci\'{o}n,Casilla 160-C, Concepci\'{o}n, Chile\\
$^{15}$NAOC-UKZN Computational Astrophysics Centre, University of KwaZulu-Natal, Durban 4000, South Africa\\
$^{16}$Kavli Institute for Astronomy and Astrophysics, Peking University, Haidian District, Beijing 100871, People’s Republic of China\\
$^{17}$Department of Astronomy, School of Physics, Peking University, Beijing 100871, People’s Republic of China\\
$^{18}$College of Science, Yunnan Agricultural University, Kunming 650201, People’s Republic of China\\
$^{19}$Institute of Astronomy and Astrophysics, Anqing Normal University, Anqing, 246133, PR China\\
$^{20}$Nobeyama Radio Observatory, National Astronomical Observatory of Japan, National Institutes of Natural Sciences, Nobeyama,\\ Minamimaki, Minamisaku, Nagano 384-1305, Japan\\
$^{21}$Department of Astronomical Science, The Graduate University for Advanced Studies, SOKENDAI, 2-21-1 Osawa, Mitaka, Tokyo 181-8588, Japan\\
$^{22}$S. N. Bose National Centre for Basic Sciences JD Block, Sector-III, Salt Lake City, Kolkata - 700 106, India\\
$^{23}$School of Physics and Astronomy, Sun Yat-sen University, 2 Daxue Road, Tangjia, Zhuhai, Guangdong Province, China\\
$^{24}$CSST Science Center for the Guangdong-Hongkong-Macau Greater Bay Area, Sun Yat-Sen University, Guangdong Province, China\\
$^{25}$Laboratory for Space Research, The University of Hong Kong, Hong Kong, China\\
$^{26}$Astronomy Department, University of California, Berkeley, CA 94720\\
$^{27}$Physical Research Laboratory, Navrangpura, Ahmedabad - 380 009, India\\
$^{28}$Indian Institute of Space Science and Technology, Thiruvananthapuram 695 547, Kerala, India\\
$^{29}$Korea Astronomy and Space Science Institute, 776 Daedeokdae-ro, Yuseong-gu, Daejeon 34055, Republic of Korea\\
$^{31}$Indian Institute of Science Education and Research (IISER) Tirupati, Rami Reddy Nagar, Karakambadi Road, Mangalam (P.O.), Tirupati 517 507, India\\
$^{33}$University of Science and Technology, Korea (UST), 217 Gajeong-ro, Yuseong-gu, Daejeon 34113, Republic of Korea\\
$^{34}$Center for Astrophysics, Harvard Smithsonian, 60 Garden Street, Cambridge, MA 02138, USA\\
$^{35}$IRAP, Université de Toulouse, CNRS, UPS, CNES, 31400, Toulouse, France

\end{document}